\documentclass[pra,aps,superscriptaddress,twocolumn]{revtex4}%
\usepackage{color} 
\usepackage{amssymb,amscd,amsmath} 
\usepackage{graphicx}
\usepackage[english]{babel}
\usepackage{epstopdf}
\usepackage{mathrsfs}
\usepackage{textcomp}
\usepackage{amsfonts}
\usepackage{amssymb}
\usepackage{hyperref}
\usepackage{bm,bbold,dsfont}
\usepackage{color}
\usepackage{xcolor}
\usepackage{times}
\usepackage[normal]{subfigure}
\usepackage{float}
\usepackage{setspace}
\usepackage{dcolumn}
\usepackage{epstopdf,epsfig}
\usepackage{makeidx}
\usepackage{tikz}

\newcommand{\change}[1]{\textcolor{black}{#1}}

\DeclareMathOperator{\sgn}{sign}

\begin{document}
\title{Extended Kitaev chain with longer-range hopping and pairing}
\author{Antonio Alecce}
\affiliation{Dipartimento di Fisica e Astronomia "G. Galilei", Universit\`a 
di Padova, via Marzolo 8, 35131  
Padova, Italy}
\author{Luca Dell'Anna\email[correspondence at: ]{luca.dellanna@unipd.it}}
\affiliation{Dipartimento di Fisica e Astronomia "G. Galilei", Universit\`a 
di Padova, via Marzolo 8, 35131  
Padova, Italy}
\affiliation
{CNISM, Sezione di Padova, Italy}

\begin{abstract}
We consider the Kitaev chain model with finite and infinite range in the hopping and pairing parameters, looking in particular at the appearance of Majorana zero energy modes and massive edge modes. We study the system both in the presence and in the absence of time reversal symmetry, by means of topological invariants and  exact diagonalization, disclosing very rich phase diagrams. In particular, for extended hopping and pairing terms, we can get as many Majorana modes at each end of the chain as the neighbors involved in the couplings. Finally we generalize the transfer matrix approach useful to calculate the zero-energy Majorana states at the edges for a generic number of coupled neighbors. 
\end{abstract}
\maketitle

\section{Introduction}
\label{sec:int}

Topological orders and their related quantum phenomena have attracted a lot of attention in the last few years \cite{Wen}. 
One of the most interesting model which shows a topological order is the Kitaev chain \cite{Kitaev}. This model describes a p-wave superconducting wire that, under certain conditions, has a gapped bulk together with zero energy 
unpaired Majorana modes localized at the edges of the system, 
which are robust against disorder \cite{Chen}, local impurities 
\cite{Ising} or dynamical perturbations \cite{DeGottardi2,Zvyagin}. 
The presence of such modes has recently been observed experimentally 
\cite{ExperimentalMFs,Exp2} and optical implementations of generalized Kitaev 
model have been proposed \cite{optics}. This one-dimensional 
system represents the simpliest playground for 
designing topological quantum computers \cite{Alicea,Akhmerov}, because of 
its straightforward mapping to a spin system \cite{Ising,Pfeuty,Lieb}.

An extension of the Kiteav model has been recently proposed by considering a 
longer range in the hopping and pairing terms appearing in the 
Hamiltonian 
\cite{Ising,DeGottardi,Ghazaryan,Lepori,Lepori2,delgado,Lepori3,Shi,Pientkaprb,Pientka,pachos}. 
We will consider such a situation 
for both long-range and finite-range interactions. 
In the latter case, if time reversal symmetry (TRS) holds, one can 
get many Majorana zero modes (MZMs) per edge as in the case of multiband spin-orbit coupled superconductor nanowire with parallel Zeeman splitting 
\cite{Tewari}. 
In this case, according to standard symmetry classification \cite{classification}, the system belongs to the class BDI, and 
the number of zero modes is dictated by the $\mathbb{Z}$-valued topological 
invariant (TI) winding number ${\sf w}$ \cite{Ising,Zinvariant1}. 
Interestingly, for certain range of parameters, the number of Majorana modes 
can be equal to the number of neighbors.   
If time reversal symmetry is broken, the system belongs to class $D$ and 
one can have at most one Majorana zero mode per edge \cite{Ising} in the regime of parameters where the pfaffian $\mathbb{Z}_2$ topological invariant \cite{Kitaev} is not trivial. 
It has been shown that such invariant, related to the fermionic parity of the ground state, corresponds to the quantized Zak-Berry phase  
\cite{Budich}. 

In this paper we study in a systematic way the phase diagrams of the 
extended Kitaev model, considering also the long-range limit both in the 
presence and in the absence of time reversal symmetry, 
by calculating the topological invariants and performing exact 
diagonalization of the Hamiltonian. 
We also provide a detailed derivation of the transfer matrix approach in the case where a generic number of neighbors are involved in the couplings and many Majorana states are located at the edges, consistently with the topological invariant analysis.
 \section{Long-Range Hamiltonian} 
 We propose an extended Kitaev chain model taking into account $r$ neighbor interactions in the hopping and pairing terms. 
We assume both of these interactions to be algebraically decreasing with the 
distance between two different lattice sites. 
The model with infinitely long-range pairing has been studied in 
Ref.~\cite{Lepori}. This model 
shows, for certain physical regimes, Majorana zero modes and massive edge modes identified as topological massive Dirac fermions in 
Ref.~\cite{delgado}, because their energies are separated from the excited states by a finite gap even in the thermodynamic limit. \\
The fermionic Hamiltonian we will consider, which generalizes the Kitaev 
chain, 
is the following 
 \begin{eqnarray}
\label{hnostra}
 H&=&\displaystyle-\sum_{j=1}^{L}\,\mu\left(a_j^{\dagger}a_j-\frac{1}{2}\right)\\
\nonumber &+&\sum_{\ell=1}^{r}\change{\sum_{j=1}^{L-\ell}}\left(-w_\ell
\, e^{i\varphi_\ell} 
a_j^{\dagger} a_{j+\ell}+\Delta_\ell\,
a_ja_{j+\ell}+\text{h.c.}\right)
 \end{eqnarray}
where $L$ is the number of lattice sites and  
$\mu$ the chemical potential. The extended hopping and pairing coupling terms, $w_\ell$ and $\Delta_\ell$, can be generic variables, although, in what follows, we will assume the following form for those parameters
\begin{equation}
\label{parameters}
w_\ell={w_0}\,{d_\ell^{-\alpha}},\;\;\;\;\Delta_\ell={\Delta}\,{d_\ell^{-\beta}},
\end{equation}
which couple the lattice site $j$ with the site $(j+\ell)$. 
The hopping parameter $w_\ell$ can acquire a phase, $w_\ell e^{i\varphi_\ell}$, with $\varphi_\ell$, $w_0$, $\Delta$ real values, in the case of broken time reversal symmetry. 
The index $\ell$ runs over the neighbor sites and $d_\ell$ is
$ d_\ell=\min(\ell,L-\ell)$ 
for closed boundary conditions and $d_\ell=\ell$ for open boundary conditions. 
The exponents $\alpha$ and $\beta$ characterize the rate of decay for the parameters if they are assumed not negative. For $\alpha\rightarrow \infty$, $\beta\rightarrow \infty$ one recovers the standard short-range Kitaev model. 
Let us suppose now to close the chain with periodic (PBC) or antiperiodic boundary conditions (ABC) and make the Fourier transform
$a_j=\frac{1}{\sqrt{L}}\sum_{k}a_{k}e^{-ik j}$
 with $k=\frac{2\pi n}{L}$ for PBC and $k=\frac{2\pi n+\pi}{L}$ for ABC,
with $n$ integer numbers. The Hamiltonian can be written as follows
 \begin{eqnarray}
\nonumber
 H&=&\sum_{k}\begin{pmatrix}
 a_{k}^{\dagger} & a_{-k}
 \end{pmatrix}\Bigg\{\sum_{\ell}\left[w_{\ell}\sin(\varphi_\ell)\sin(k \ell)\right]\mathds{1}\\
 \nonumber
&&-\left(\frac{\mu}{2}+\sum_{\ell}\left[w_{\ell}\cos(\varphi_\ell)\cos(k \ell)\right]\right){\sigma_z}\\
\nonumber 
&&+\left(\sum_\ell \Delta_\ell\sin(k \ell)\right){\sigma_y}\Bigg\}
\begin{pmatrix}
 a_{k}\\
 a_{-k}^{\dagger}
 \end{pmatrix}\\
 &\equiv&\sum_{k}\begin{pmatrix}
 a_{k}^{\dagger} & a_{-k}
 \end{pmatrix}{\cal H}(k)\begin{pmatrix}
 a_{k}\\
 a_{-k}^{\dagger}
 \end{pmatrix}
\label{kspace}
 \end{eqnarray}
The spectrum has a particle-hole symmetry since, 
under particle-hole transformation, ${\cal C}=\sigma_x{\cal K}$ (where ${\cal K}$ is the complex conjugation operator),   
 \begin{equation}\label{ph}
 {\cal C}^{-1}{\cal H}(k){\cal C}=-{\cal H}(-k).
 \end{equation}
The time reversal condition ${\cal H}(-k)^{*}={\cal H}(k)$ is satisfied only if we consider real hopping terms $w_\ell$, thus if $\varphi_\ell=0,n\pi$. In the next sections we will analyze both situations, with and without TRS.

Before we proceed a comment about boundary conditions is in order. 
In Ref.~\cite{Lepori}, ABC are assumed to preserve pairing terms in $H$, on the same time this choice would destroy long-range hopping terms for finite $L$ and $r\ge L/2$. For PBC the reverse holds. If $r<L/2$, instead, both choices preserve long-range interaction terms and translational invariance. 
Closed chains with ABC and PBC, however, may give different results. 
For finite $r$, in the expression for ${\cal H}(k)$, terms like 
$\cos\left(\ell\left(\frac{2\pi n}{L}+\frac{\pi}{L}\right)\right)$ and 
$\sin\left(\ell\left(\frac{2\pi n}{L}+\frac{\pi}{L}\right)\right)$ for ABC can be confused with the corresponding PBC conterparts, 
$\cos\left(\ell\left(\frac{2\pi n}{L}\right)\right)$ and $\sin\left(\ell\left(\frac{2\pi n}{L}\right)\right)$, in the limit $L\rightarrow\infty$, if $\ell$ remains finite. 
If, instead, we consider an infinite number of interacting neighbors (always in the thermodynamic limit, $L\rightarrow \infty$), $d_\ell$ is always $\ell$ and the above terms in \eqref{kspace} generates polylogarithmic functions $Li_{\alpha}(e^{\pm ik \ell})$ where $k$ becames a continuum variable. In the thermodynamic limit, therefore, ${\cal H}(k)$ is always the same, for both PBC and ABC.

\section{Topological invariants}
Topological phases are described by different topological invariants according to the presence or the absence of time reversal symmetry. 
If time reversal is preserved ($\varphi_\ell=0$) there is a $\mathbb{Z}$ topological invariant given by the winding number. 
In this case we can define the 
vector
 \begin{equation}
 \textbf{h}(k)=\left(0, h_y(k),  h_z(k)\right)
 \end{equation}
such that ${\cal H}(k)$ in Eq.~(\ref{kspace}), for $\varphi_\ell=0$, 
can be written as
\begin{equation}
{\cal H}(k)= \textbf{h}(k)\cdot\bm{\sigma}
\end{equation}
where $\bm{\sigma}$ is the vector made of Pauli matrices. In terms of unit vector $\hat{\textbf{h}}(k)=\textbf{h}(k)/|\textbf{h}(k)|$ we define the winding number 
 \begin{equation}\label{winding}
 {\sf w}=\frac{1}{2\pi}\oint\,d\theta_{k}=\frac{1}{2\pi}\int_{-\pi}^{\pi}d k\frac{\partial_{k}\hat{h}_z(k)}{\hat{h}_y(k)}
 \end{equation}
 \\
In the case of broken TRS ($\varphi_\ell\neq 0$), as we can see from Eq.~\eqref{kspace}, we also have to consider the component of ${\cal H}(k)$ proportional to $\mathds{1}$ (or by a transformation, proportional to ${\sigma}_x$) so that the winding number ${\sf w}$ cannot be defined. For $k=0,\pm\pi$ we have only the Hamiltonian term along ${\sigma}_z$, while the term along ${\sigma}_y$ vanishes. 
In such a situation a good topological invariant is given by 
 \begin{equation}
\label{upsilon}
 \upsilon= \textrm{sign}( h_z(0)\, h_z(\pi))
 \end{equation}
which can be used also for time reversal symmetry cases.
 \section{Finite number of neighbors: topological phase diagrams}
Here we will focus on the case of finite number of interacting neighbors. At first we will allow the hopping term to be longer ranged, keeping the pairing interaction short-ranged, and then we will consider the opposite case. At the end we will allow both terms to be longer ranged, analyzing the topological phase in the presence or in the absence of TRS.
 \subsection{Extended-range hopping}
\begin{figure}
 	\centering 
 	\subfigure[ $\,r=2,\;\alpha=0,\,\beta\rightarrow\infty$]{
 		\includegraphics[width=.75\columnwidth]{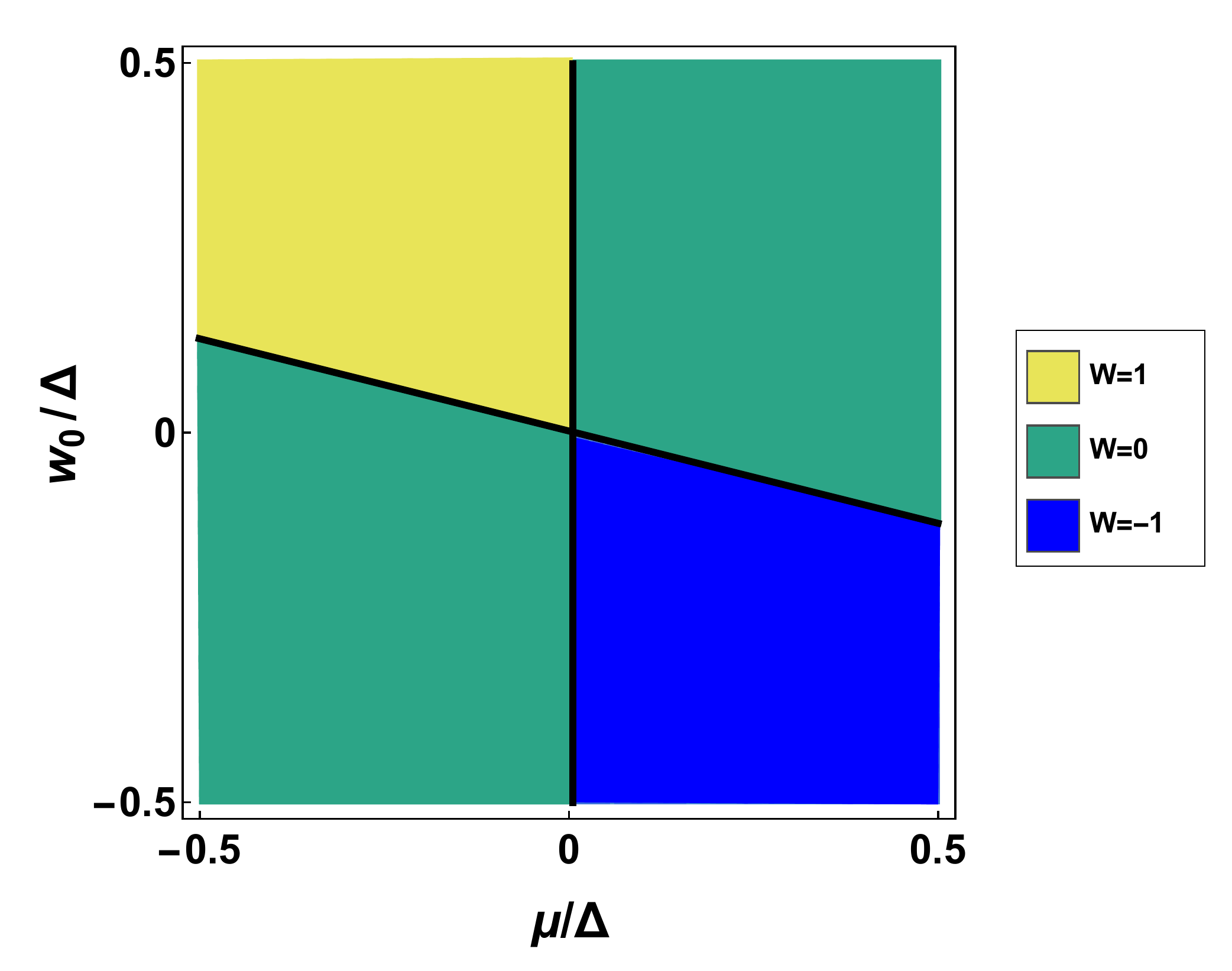}
 		\label{fig:hoppingnv2alpha0}}
 	\subfigure[ $\,r=2,\;\alpha=0.3,\,\beta\rightarrow\infty$]{
 		\includegraphics[width=.75\columnwidth]{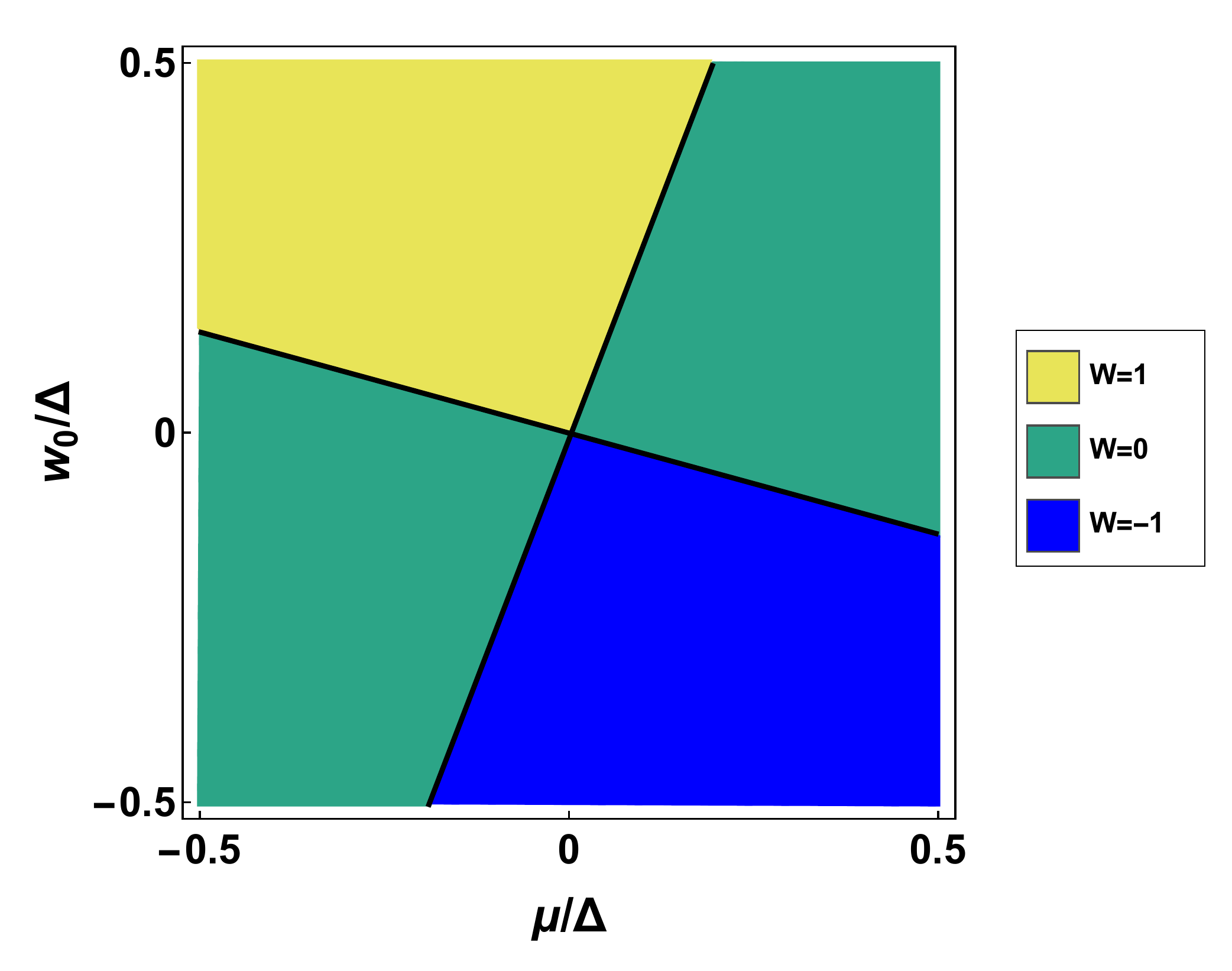}
 		\label{fig:hoppingnv2alpha03}}
 	\caption{Phase diagram ($w_0$ vs $\mu$, in units of $\Delta$), showing the values of ${\sf w}$ as in the legend, for $r=2$ for two different values of $\alpha$, in the limit of $\beta\rightarrow\infty$, i.e. only extended hopping. Topological regime increases with $\alpha$, and saturates for large $\alpha$ to the topological phase of the standar Kitaev model defined by the condition $-2 < \mu/w_0 < 2$.}
 	\label{fig:solohopping}
 \end{figure}

\begin{figure}
        \centering
        \subfigure[ $\;r=3, \;\alpha,\,\beta\rightarrow\infty$ ]{
                \includegraphics[width=0.45\columnwidth]{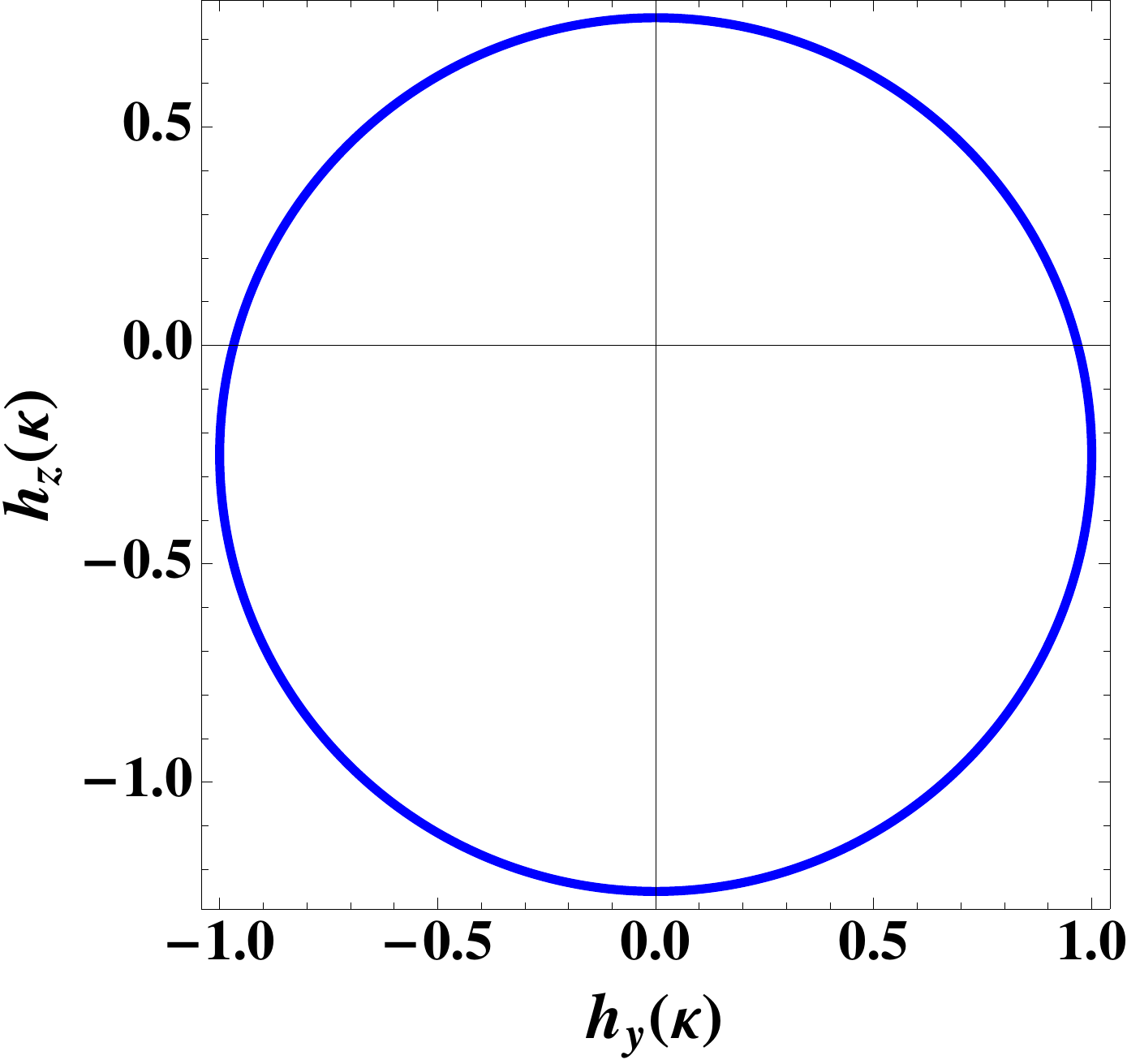}
                \label{fig:o}}
 	\centering
 	\subfigure[ $\;r=3, \;\alpha=0,\,\beta\rightarrow\infty$ ]{
 		\includegraphics[width=0.45\columnwidth]{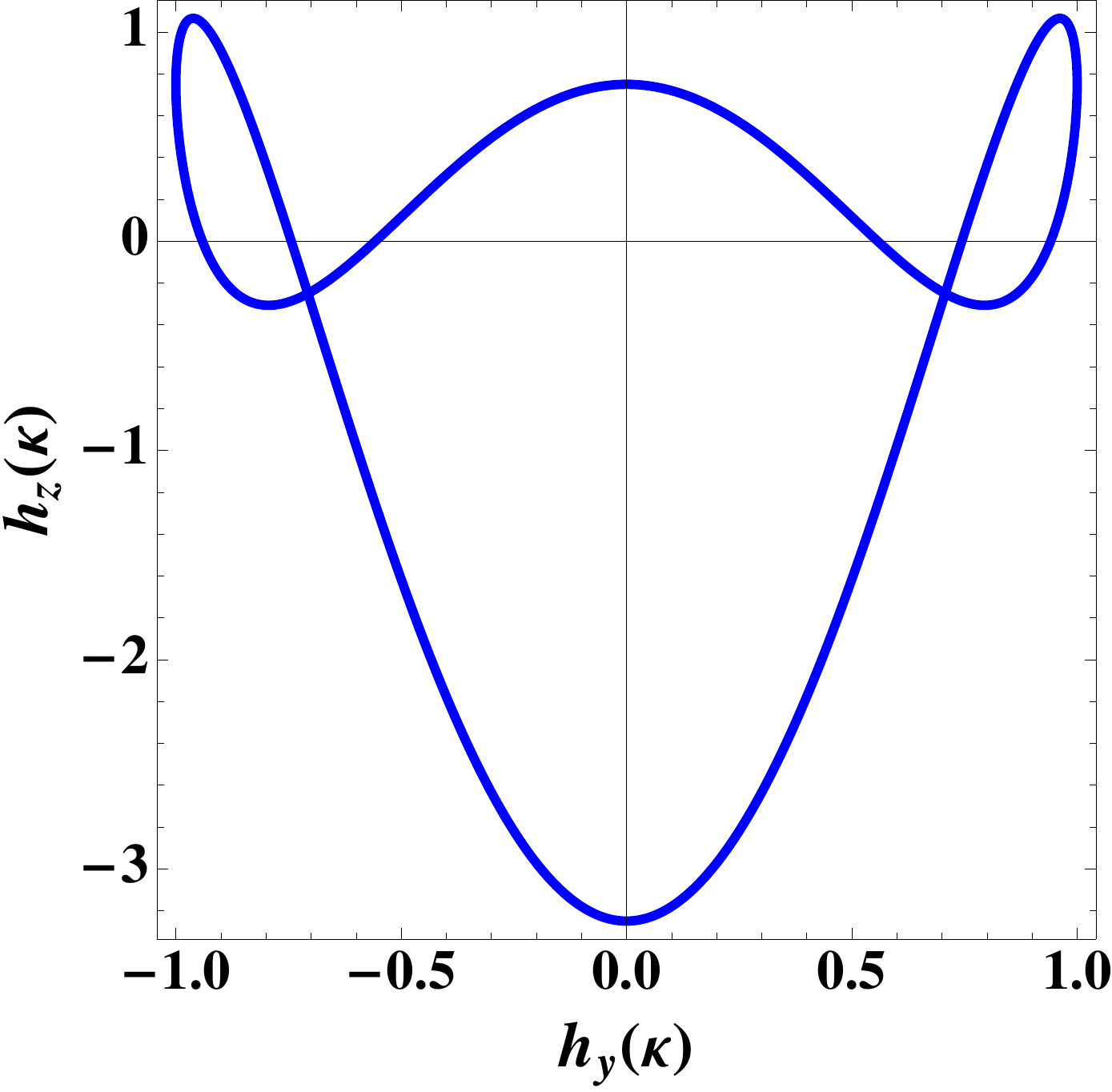}
 		\label{fig:yzsolohopping}}
        \centering
        \subfigure[ $\;r=3, \;\beta=0,\,\alpha\rightarrow\infty$]{
                \includegraphics[width=0.45\columnwidth]{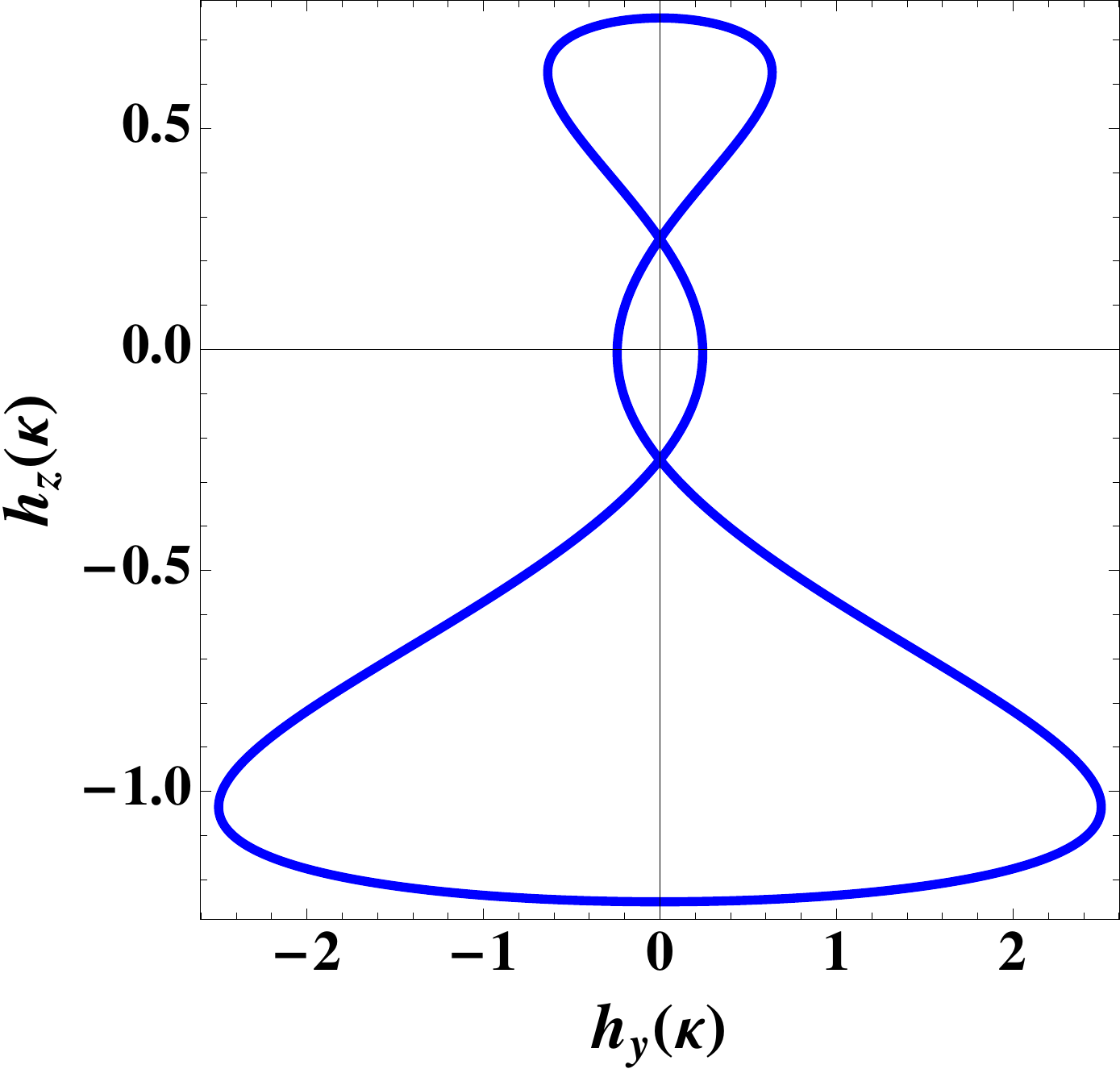}
                \label{fig:2c}}
 	\centering
 	\subfigure[ $\;r=3,\;\alpha=\beta=0$]{
 		\includegraphics[width=0.45\columnwidth]{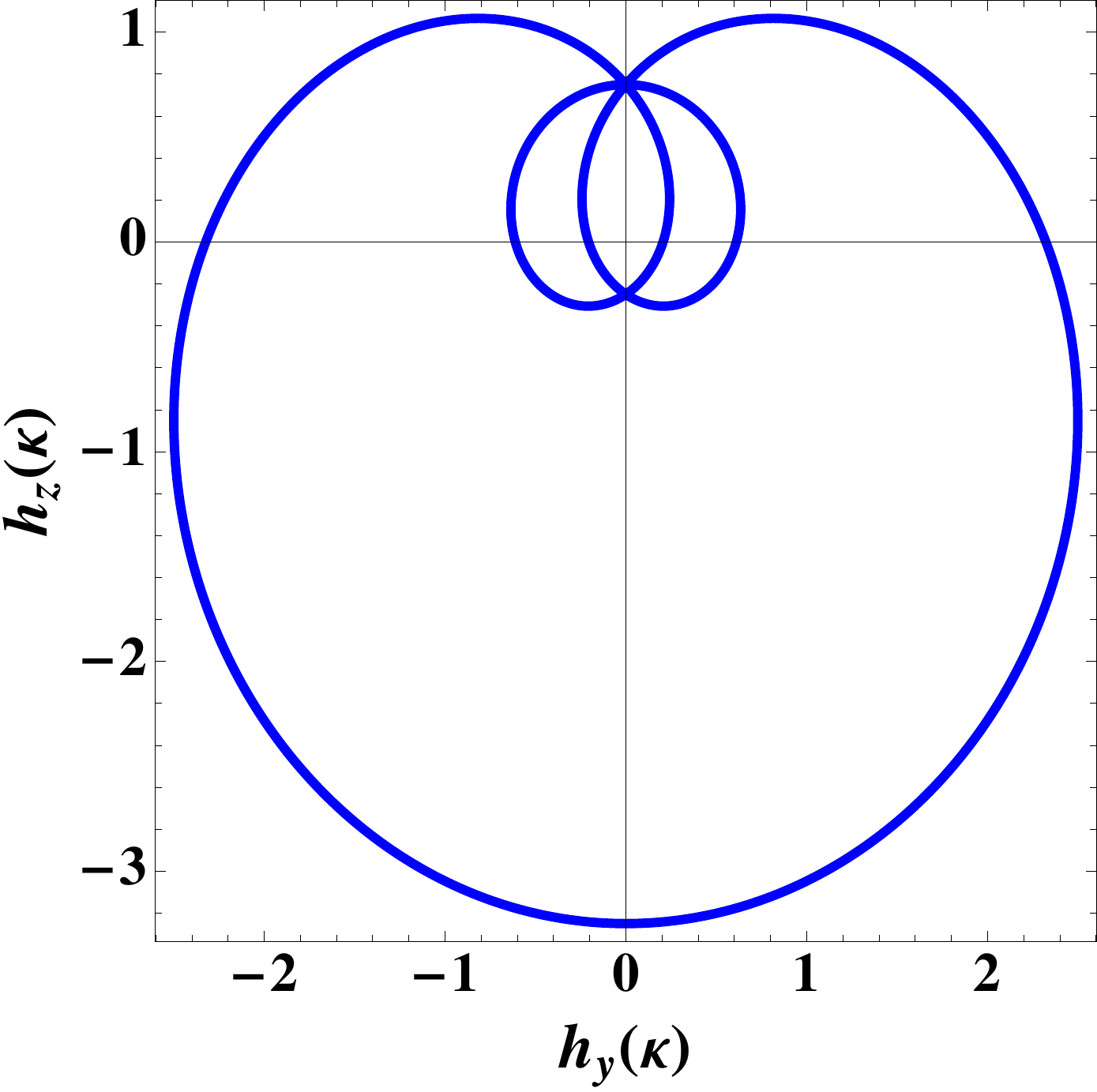}
 		\label{fig:yzr3}}
 	\caption{Winding vector ${\bf h}(k)$ (not normalized) with $r=3$, for 
(a) 
$\alpha,\beta\rightarrow\infty$ (short range in the hopping and pairing), (b) $\alpha=0$ and $\beta 
\rightarrow\infty$ (short range in the pairing and extended range in the hopping),  
(c) $\alpha\rightarrow\infty$, $\beta=0$ (short range in the hopping and extended range in the pairing) 
and (d) $\alpha=\beta=0$ (extended hopping and pairing). 
In all the figures $\mu/\Delta=0.5$ and $w_0/\Delta=1$. In the first three cases going along the path one goes around $(0,0)$ only once, in the last case one can make $1$ or $3$ twists around the origin.}
 	\label{fig:yzplane}
 \end{figure}

 We will now analyze Eq.~\eqref{hnostra} in the limit of only extended-range hopping, i.e. $\beta\rightarrow\infty$, for finite $r$. 
 In Fig.~\ref{fig:solohopping} we observe that the topological regime (${\sf w}=\pm1$) of parameters increases with $\alpha$, the exponent of the hopping 
term. The result we found for this specific Hamiltonian is consistent with the one obtained in \cite{delgado}. Only one MZM per edge can be found, since the dependence of the pairing term on $k$ goes as $\sin(k)$ and not as $\sin(\ell k)$ ($\ell=1,\dots,r$). This means that moving along the first Brillouin zone (in $k$-space), the winding vector makes only one cycle around $(0,0)$. 
To better show this aspect we report in Fig.~\ref{fig:yzsolohopping} the points spanned by the unormalized winding vector $\textbf{h}(k)=\left(0,\Delta\sin(k),-\mu/2-w_0\sum_\ell \ell^{-\beta}\cos(\ell k)\right)$. 
At any finite number of neighbors $r\le L/2$ and exponent $\alpha$, the topological regime for $w_0>0$ is defined by the condition
\begin{equation}
\label{condhop}
-2 w_0 \,H_r^{\alpha} < \mu < 2 w_0 \,{R}_r^{\alpha}  
\end{equation}
where
\begin{eqnarray}
&&H_r^{\alpha}=\sum_{\ell=1}^r {1}/{\ell^{\alpha}}\\
&&R_r^{\alpha}=\sum_{\ell=1}^r(-1)^{\ell+1}/\ell^{\alpha}
\end{eqnarray}
Here $H_r^{\alpha}$ is the generalized harmonic number of order $\alpha$ of 
$r$, and $R_r^{\alpha}=(1-2^{1-\alpha})\zeta(\alpha)-\zeta(\alpha,1+r)+2^{1-\alpha}\zeta(\alpha,1+\lfloor r/2\rfloor)$, 
with $\zeta(\alpha,n)=\sum_{\ell=0}^\infty 1/(\ell+n)^{\alpha}$ the generalized Riemann zeta function.  
For $\alpha\rightarrow \infty$ (short-range limit), $H_r^{\infty}=R_r^{\infty}=1$, and Eq.~(\ref{condhop}) reduces to the standard Kitaev condition. 
In the case of infinite number of neighbors ($r\rightarrow \infty$), in the thermodynamic limit, we get $H_\infty^{\alpha}=\zeta(\alpha)$, the Riemann zeta 
function, and $R_\infty^{\alpha}=\eta(\alpha)=(1-2^{1-a})\zeta(\alpha)$ the Dirichlet eta function.  
 \subsection{Extended-range pairing}
\begin{figure}
 	\centering 
 	\subfigure[$\,r=3,\;\beta=0,\,\alpha\rightarrow\infty$]{
 		\includegraphics[width=0.75\columnwidth]{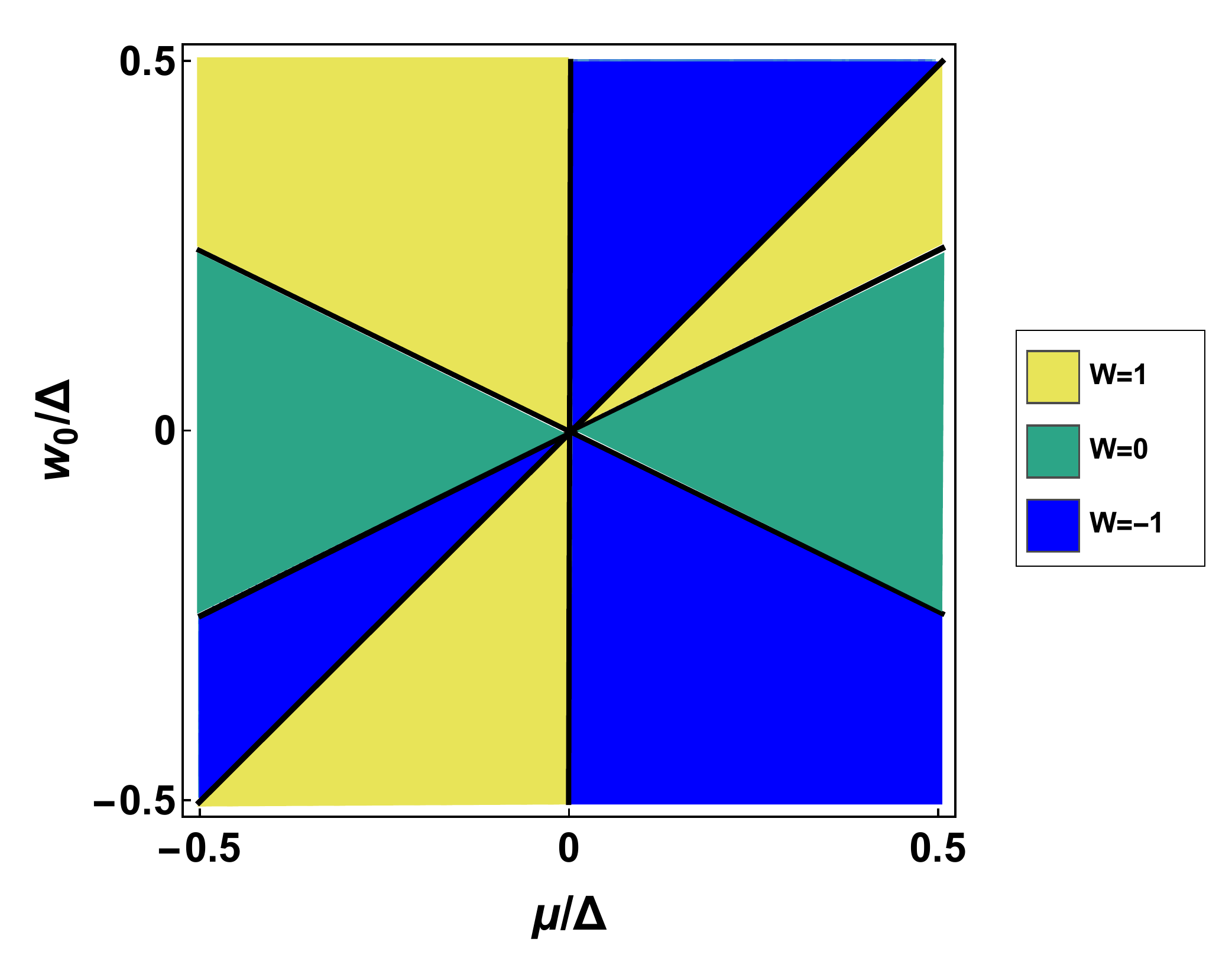}
 		\label{fig:pairingnv3beta0}}
 	\caption{Phase diagram for only extended pairing. The values of ${\sf w}$ are reported in the legend. We note in the diagram an alternating behavior of the winding number. However the regime where one Majorana mode per edge appears is the same as in the standard short-range Kitaev chain model, $-2 < \mu/w_0< 2$. }
 	\label{fig:solopairing}
 \end{figure}

 We now analyze longer range Hamiltonian with a finite number $r$ of neighbors only in the pairing term, namely $\alpha\rightarrow\infty$ and choosing a finite $\beta$. Again the winding number, as before, can take only the values 
$0,\pm 1$, however, we observe an alternation, inside the topological phase, of the values $\pm1$ (see Fig.~\ref{fig:solopairing}). 
This behavior is related to the nodes along $h_y(k)=\Delta\sum_{\ell=1}^r\change{\sin(k\ell)/{\ell}^\beta}=0$, as shown in Fig.~\ref{fig:2c}, so that the origin can be surrounded clock- or 
counterclockwise. Notice that the full size of the topological phase is the same as that of the short-range model. 
This behavior, not present in the extended hopping Hamiltonian, is more emphatic as $r$ increases (we are always assuming $r$ to be finite) while it disappears as $\beta\rightarrow\infty$, in which case we recover the topological phase diagram of the usual Kitaev model \cite{Kitaev}. 
 \subsection{Extended-range hopping and pairing, with and without TR symmetry}
 We will look now to the topological phases when considering longer range in the hopping and in the pairing at the same time, assuming $\beta=\alpha$, in the presence or in the absence of time reversal symmetry.

  \subsubsection{With time reversal symmetry}

\begin{figure}[h]
 	\centering 
 	\subfigure[$\,r=2,\;\alpha=\beta=0$]{
 		\includegraphics[width=.75\columnwidth]{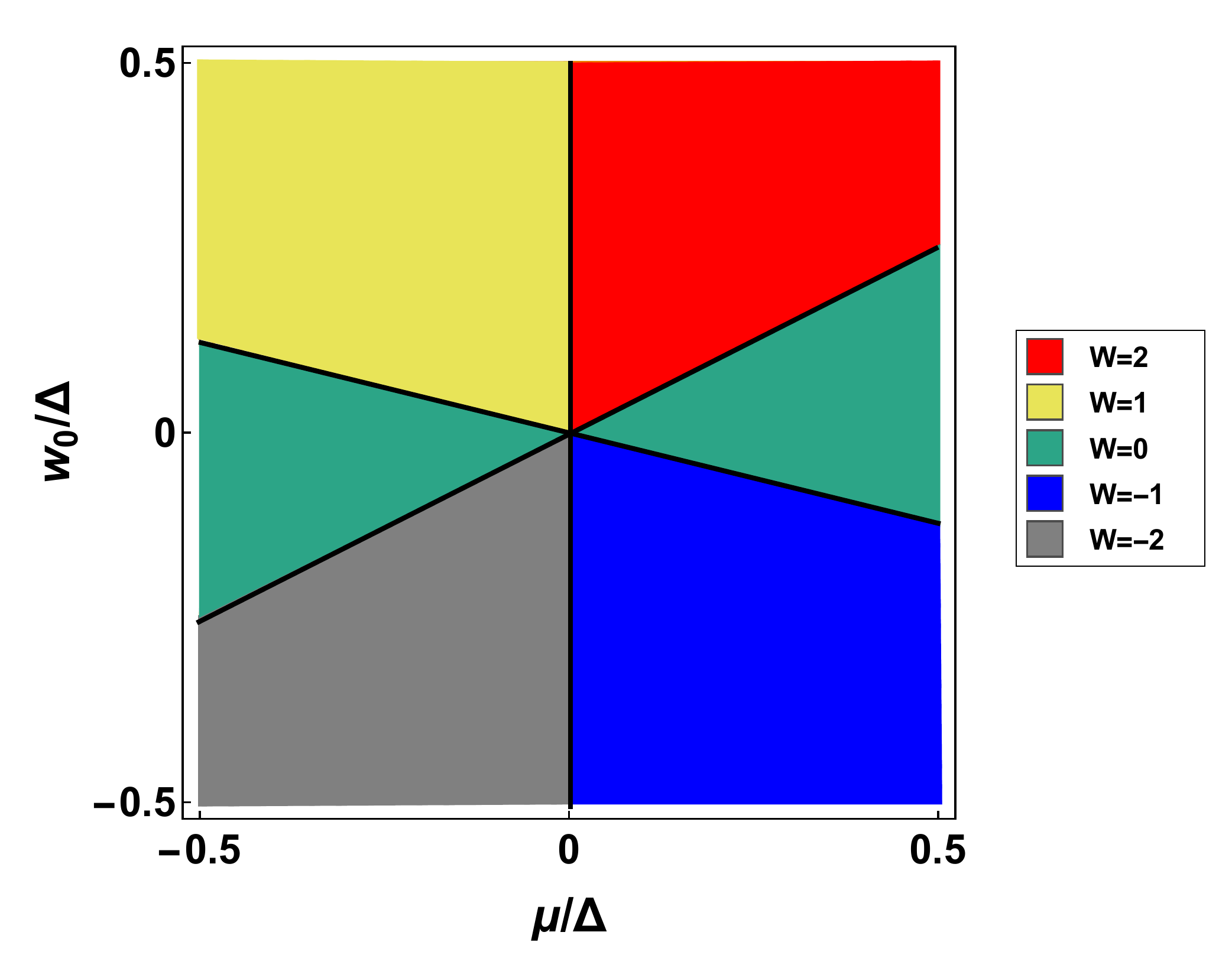}
 		\label{fig:alpha0}}
 	\subfigure[$\,r=2,\;\alpha=\beta=0.5$]{
 		\includegraphics[width=.75\columnwidth]{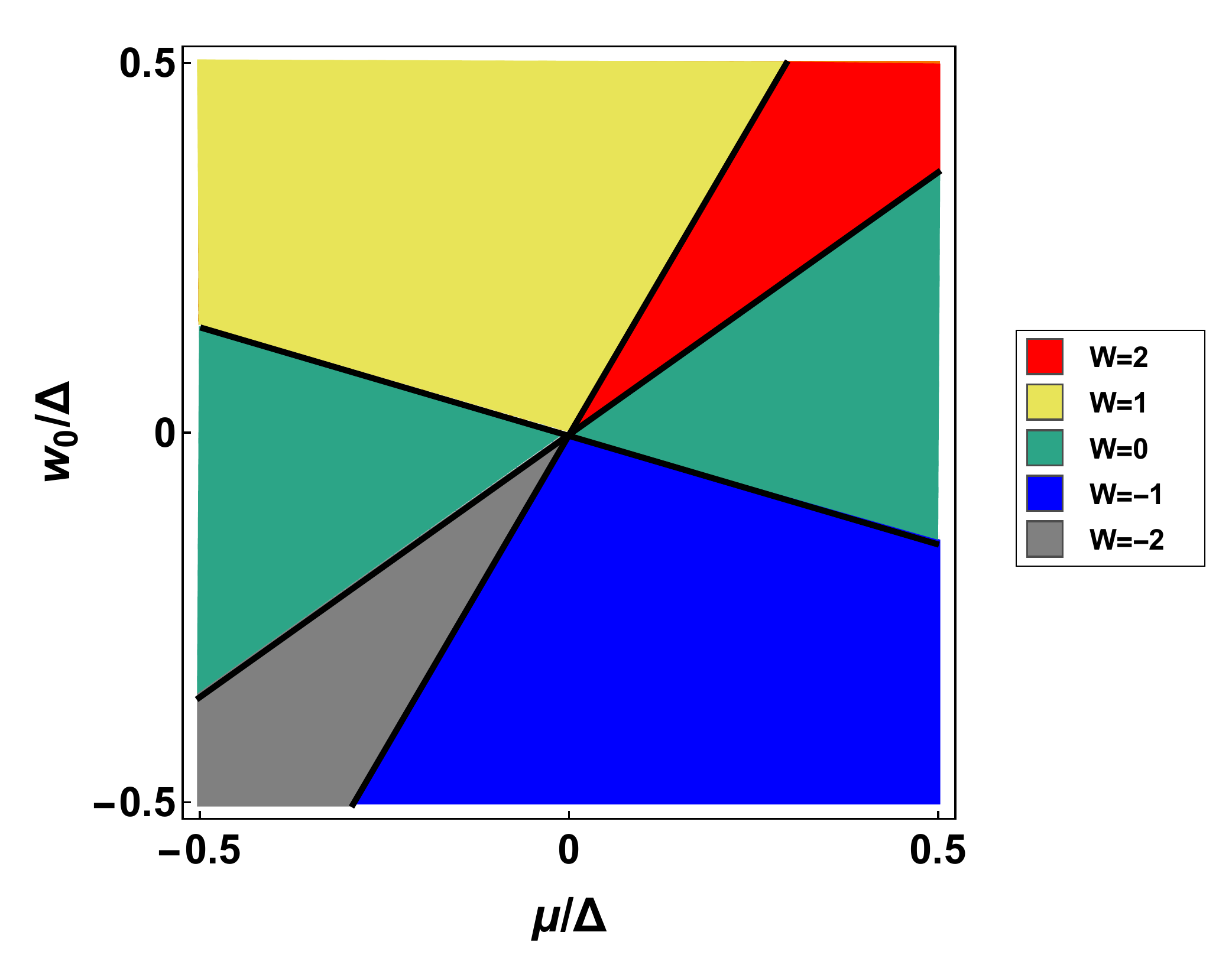}
 		\label{fig:alpha0.5}}
 	\subfigure[$\,r=2,\;\alpha=\beta=1$]{
 		\includegraphics[width=.75\columnwidth]{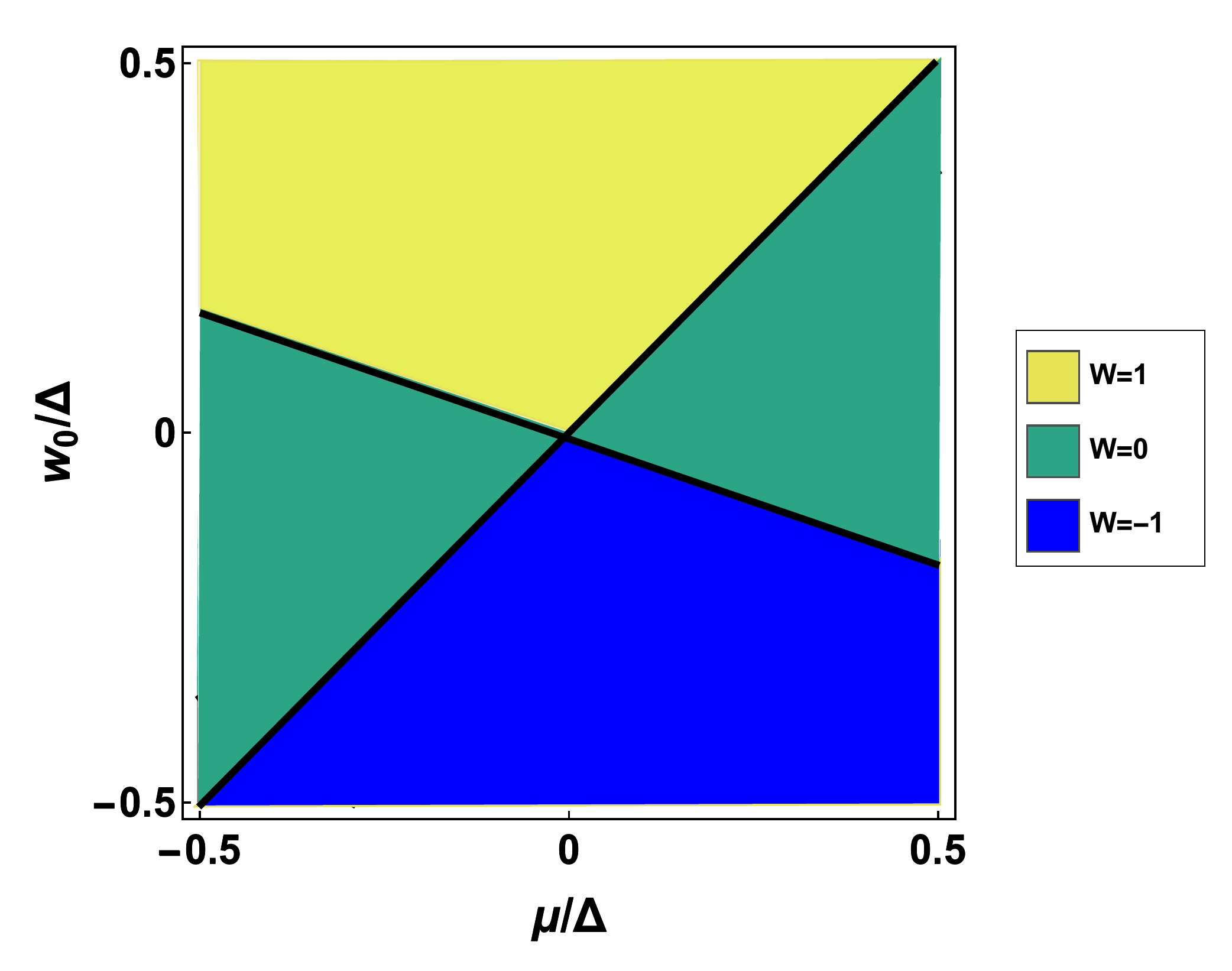}
 		\label{fig:alpha1}}
 	\caption{Phase diagram for two-neighbor hopping and pairing with $\alpha=\beta$. Only for $\alpha<1$ we can have two MZM per edge in ranges of parameters which decreases by increasing $\alpha$. }
 	\label{fig:r2}
 \end{figure}

\begin{figure}[h]
 	\centering 
        \includegraphics[width=0.9\columnwidth]{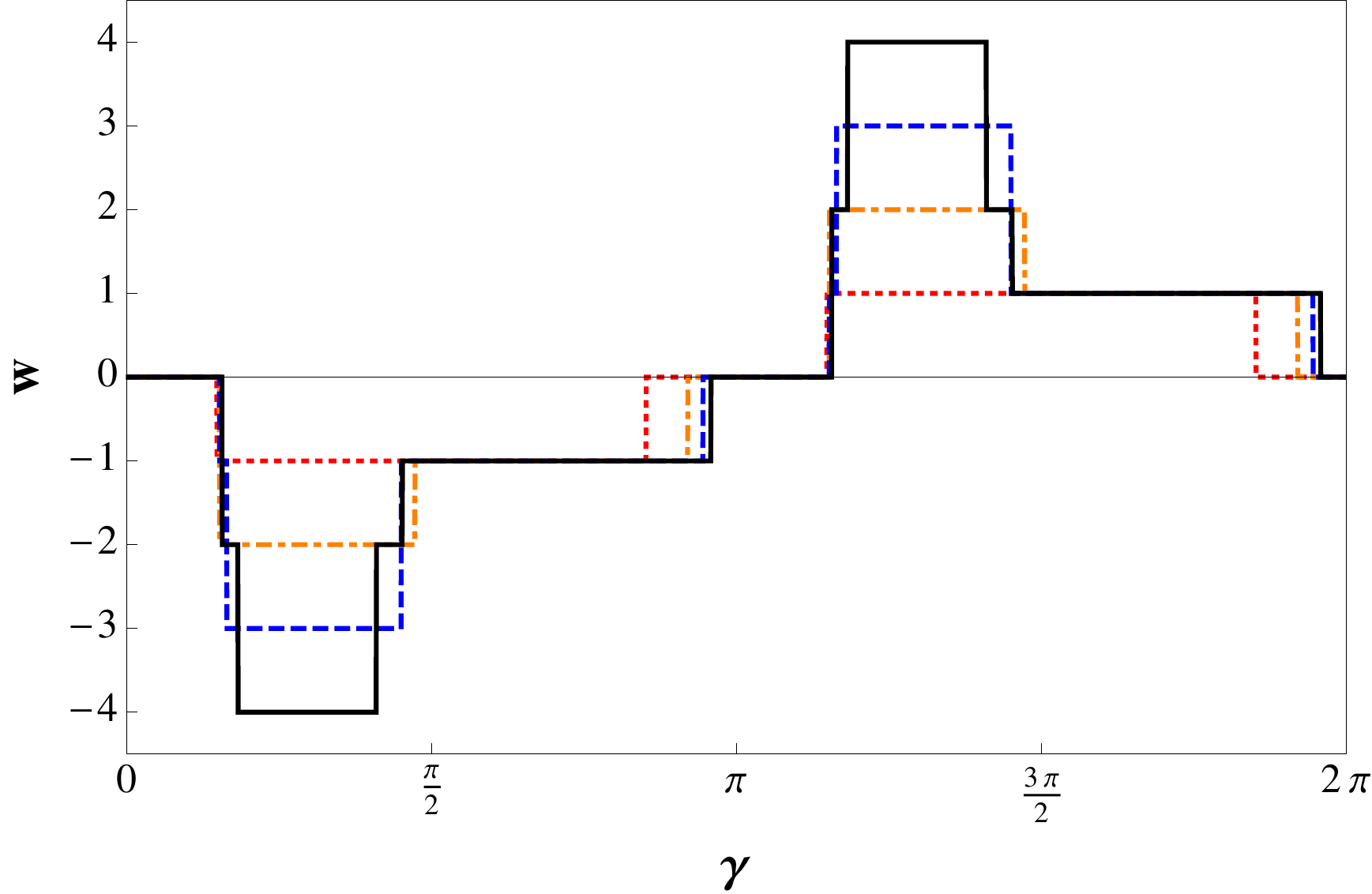} 
\caption{Values of the winding number ${\sf w}$ as a function of $\gamma$, 
an angle which parametrizes $\mu$ and $\Delta$ as it follows: $\mu/\Delta=\cos\gamma$ and $w_0/\Delta=\sin\gamma$. The numbers of neighbors 
involved in the hopping and in the pairing are $r=1$ (red dotted line), $r=2$ (dot-dashed orange line), $r=3$ (dashed blue line), $r=4$ (black solid line). 
In all the cases $\alpha=\beta=0.05$. For some values of $\gamma$ we get ${\sf w}=r$.}
 	\label{fig:5}
 \end{figure}

After considering separately the extended range in the hopping and in the pairing, we now include them together. 
The winding numbers which characterize the topological phases provide a way to count the edge modes 
of the chain in the open configuration \cite{readegreen}. In the special case where both 
the hopping and the pairing terms involve many neighbors ($r\ge1$) we observe that the winding number, and therefore,
the number of Majorana modes at each edge, can be larger than one, up to $r$. 
In order to prove this result, let us consider uniform case ($\alpha=\beta=0$) of $r$-neighbor 
hopping and pairing. In this case we have
\begin{eqnarray}
h_y(k)&=&\Delta\sum_{\ell=1}^r\sin(\ell k)\\
\nonumber&=&\Delta \sin\big((r+1)k/2\big) \sin(rk/2) \csc(k/2)\\
h_z(k)&=&-\mu/2-w_0\sum_{\ell=1}^r\cos(\ell k)\\
\nonumber&=&-\mu/2-w_0\cos\big({(r+1)k}/{2}\big) \sin({rk}/{2}) \csc({k}/{2})
\end{eqnarray}
The $2r$ zeros of $h_y(k)$, namely $k_n$ such that $h_y(k_n)=0$, can be ordered,
$k_0<k_1<k_2<k_3<....<k_{2r-1}$, and are 
\begin{equation}
k_{2n}=\frac{2\pi n}{r}, \;\;\;
k_{2n+1}=\frac{2\pi(n+1)}{r+1}\,.
\end{equation}
The corresponding values of $h_z(k_n)$ are
\begin{eqnarray}
\label{proof}
\nonumber&&h_z(0)=-\mu/2-r\,w_0\\
&&h_z(k_{2n})=-\mu/2\\
\nonumber&&h_z(k_{2n+1})=-\mu/2-w_0
\end{eqnarray}
In the regime of parameters where $h_z(k_n)$ has alternate signs for ordered $\{k_n\}$, varying $k$ from $0$ to $2\pi$ the winding vector surrounds the origin $r$ times. As a result, for $\alpha=\beta=0$, $w_0>0$, we get from Eq.~(\ref{proof}) the following topological phases
\begin{eqnarray}
\label{w1}
{\sf w}=1, \;\;&\textrm{for}&\;\;-2\, r < \mu/w_0 \le 0,\\
{\sf w}=r, \;\;&\textrm{for}& \;\;\;\;\;0 < \mu/w_0 < 2,
\label{wr}
\end{eqnarray}
and ${\sf w}=0$ otherwise. This result is verified numerically in 
Fig.~\ref{fig:r2} a) for $r=2$ where
 we obtain $\left|{\sf w}_{max}\right|=2$ which should correspond to having $2$ MZM per edge \cite{readegreen}. For a generic value of $\alpha$ the calculation is more involved, but we observe that by increasing $\alpha$, the range of parameters in which ${\sf w}$ is maximum decreases and for $\alpha\rightarrow\infty$ we recover the phase diagram of the standard Kitaev chain model. 
However, increasing $\alpha$, in some range of parameters, ${\sf w}$ can take intermediate values between $1$ and $r$ with steps of $2$, namely if $r$ is even ${\sf w}$ can be equal to $0$ (trivial phase), $1$ or any even number $\le r$, while if $r$ is odd, ${\sf w}$ can take the values $0$, $1$ or any odd number $\le r$ (see Fig.~\ref{fig:5}, where we parametrize $\mu/\Delta=\cos\gamma$ and $w_0/\Delta=\sin\gamma$ and plot ${\sf w}$ varying $\gamma$ for $\alpha=\beta=0.05$ and different values of $r$).  
This behavior is due to the mirror symmetry of the map $h_z(h_y)$ with respect to $h_y=0$, as one can see in Fig.~\ref{fig:yzr3}. This means that the edge modes are created or annihilated in pair at each edge, 
 with the only exception of the last Majorana modes. 
This behavior can be explained by the fact that increasing $\alpha$ for some range of parameters the high degeneracy at zero energy level is hardly sustained and it is easier for two Majorana modes at the same edge to annihilate each other because of the overlap. Since the system is particle-hole symmetric the same process occurs also on the other edge. If $r$ is even the last four zero modes can annihilate each other on each edge (going to the trivial phase) or by one per edge, by anti-dimerization, inceasing $\mu$. 
The last two Majorana modes, one at each edge, when ${\sf w}=1$, instead, are more robust because their annihilation requires an overlap between the two wavefunctions peaked at long (infinite) distance as in the usual Kitaev chain.


 \subsubsection{With broken time reversal symmetry}
\begin{figure}
   	\centering
   		\includegraphics[width=0.4\columnwidth]{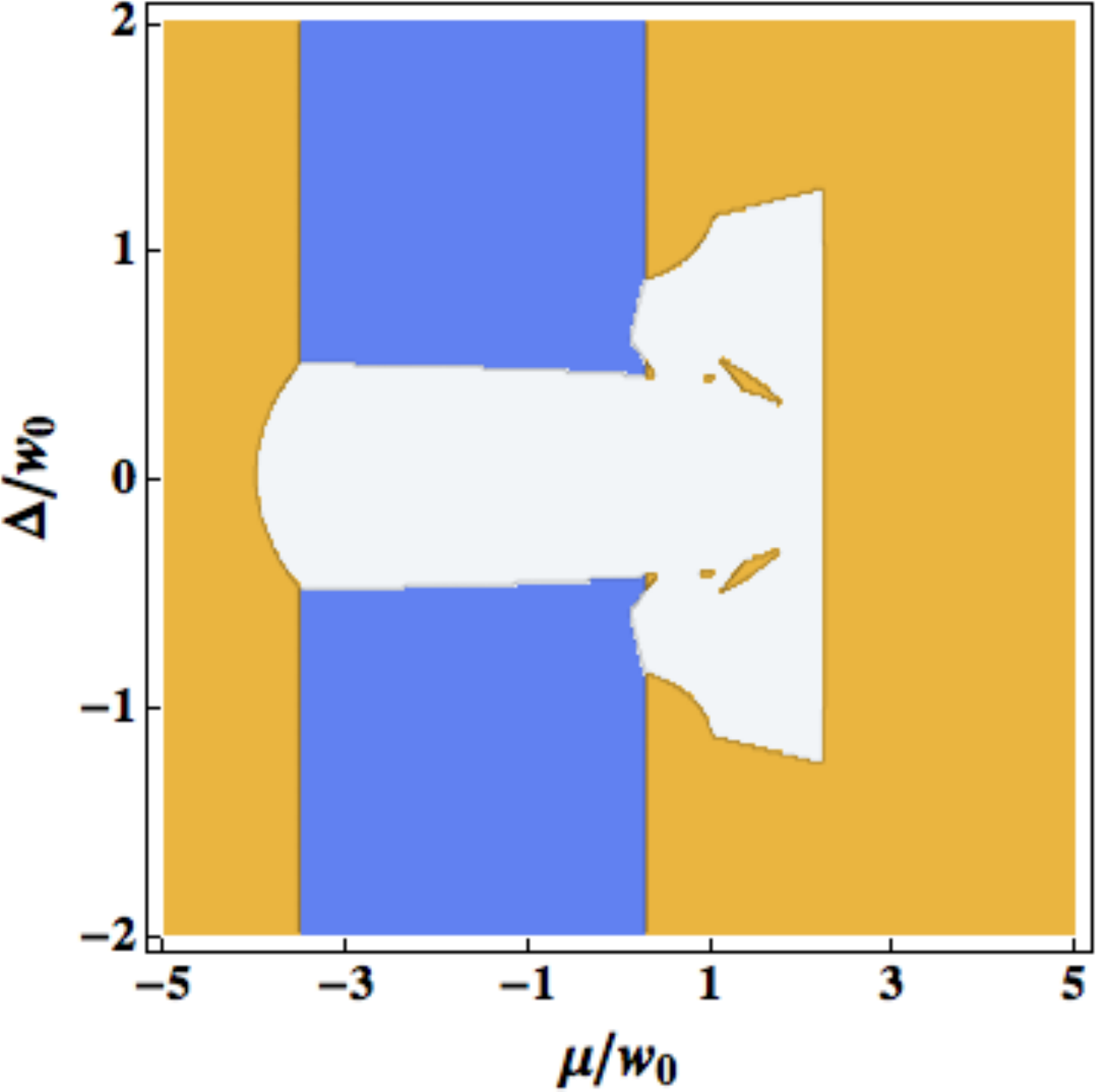}
   		\label{fig:BTRlpisu10}
   		\includegraphics[width=0.4\columnwidth]{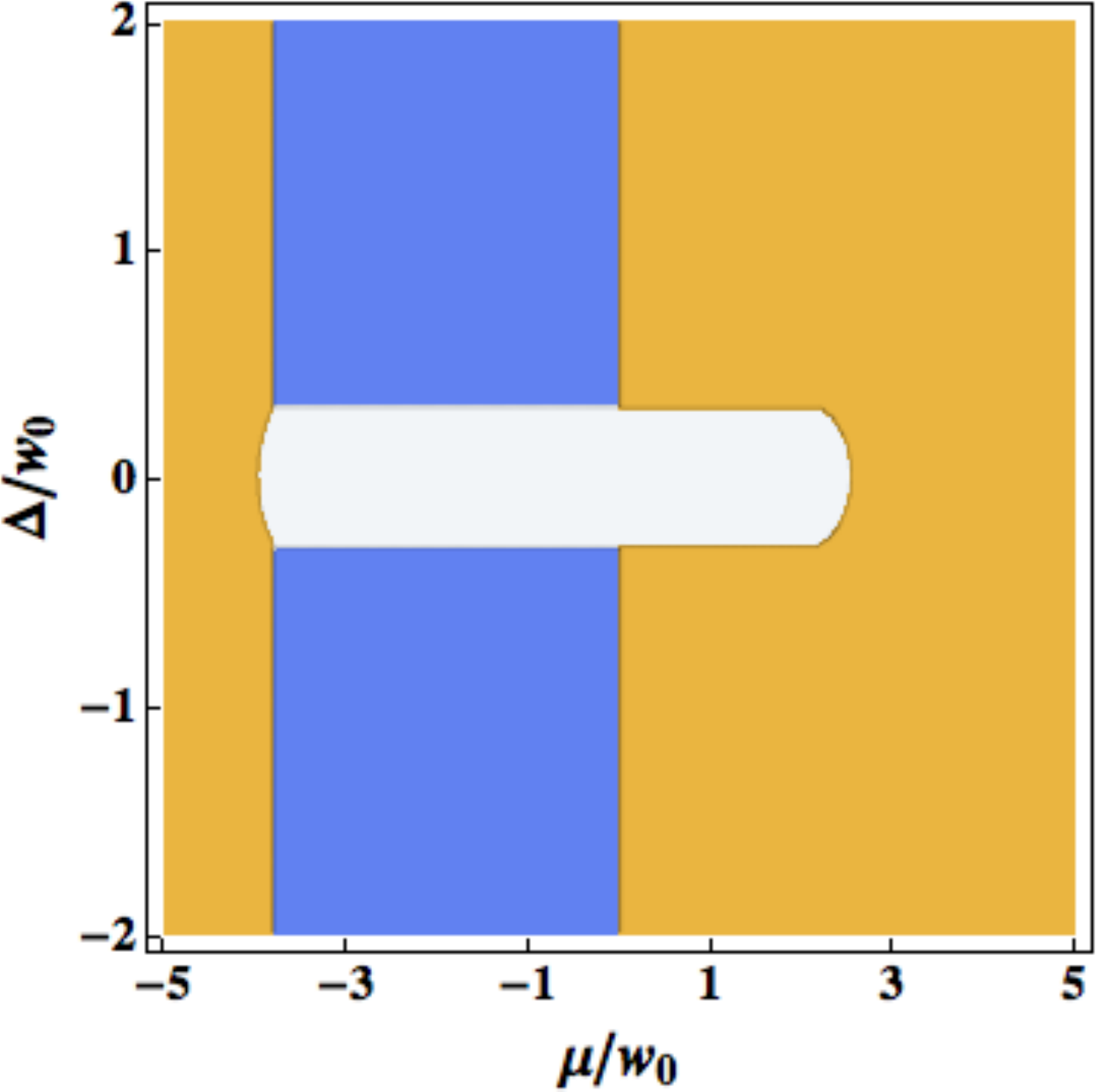}
   		\label{fig:BTRpisu10}
   		\includegraphics[width=0.4\columnwidth]{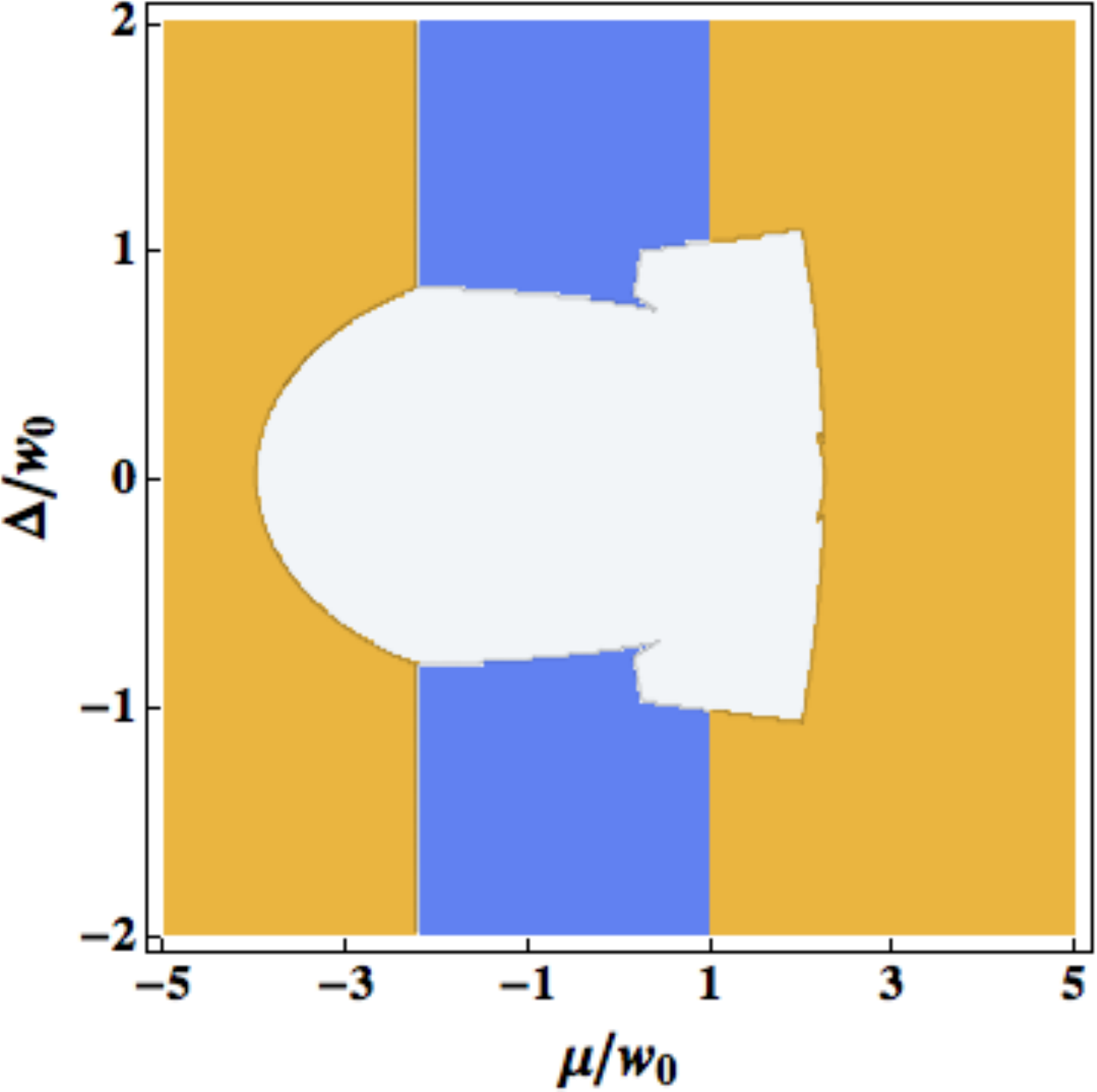}
   		\label{fig:BTRl2pisu10}
   		\includegraphics[width=0.4\columnwidth]{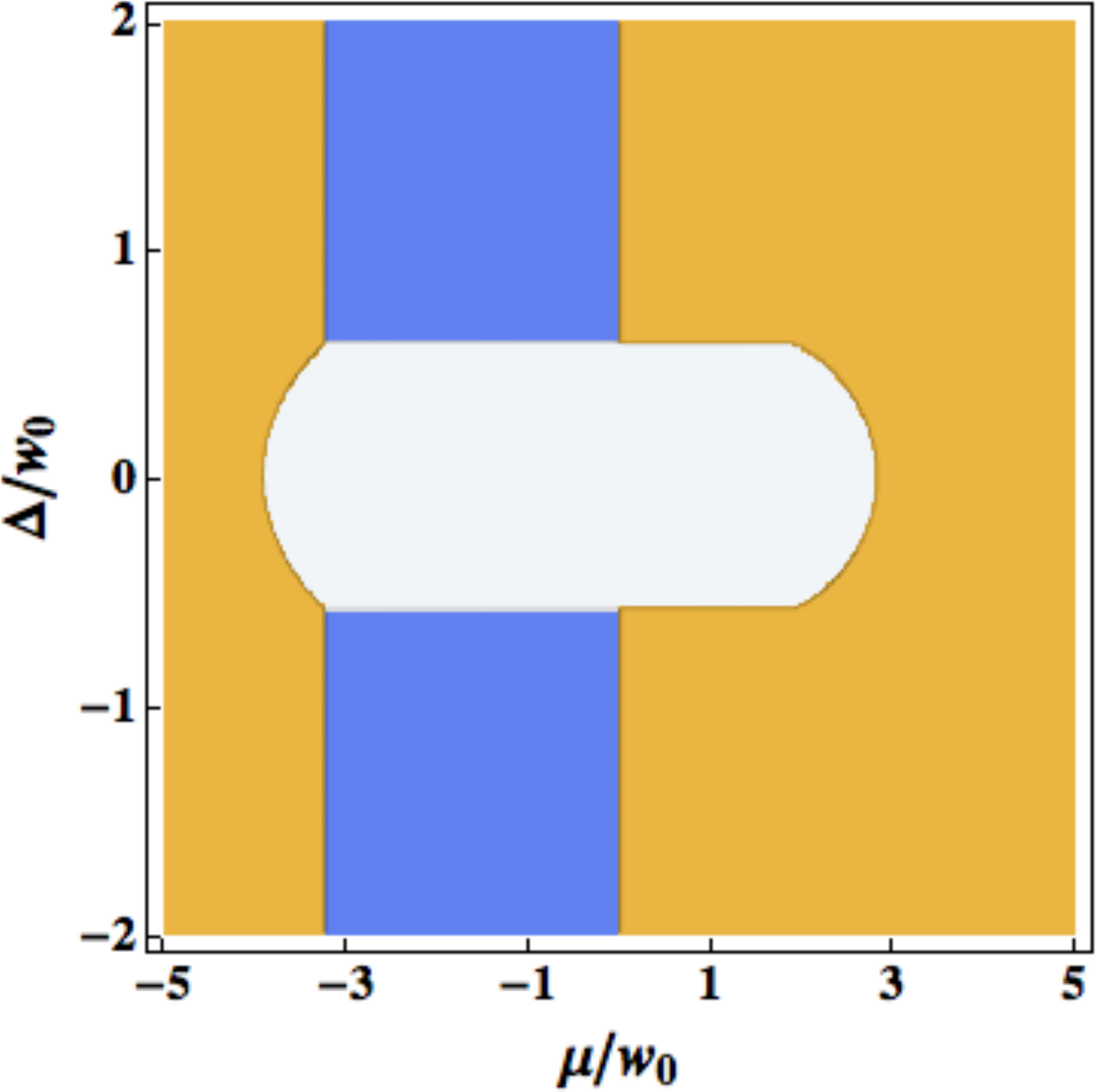}
   		\label{fig:BTR2pisu10}
                \includegraphics[width=0.4\columnwidth]{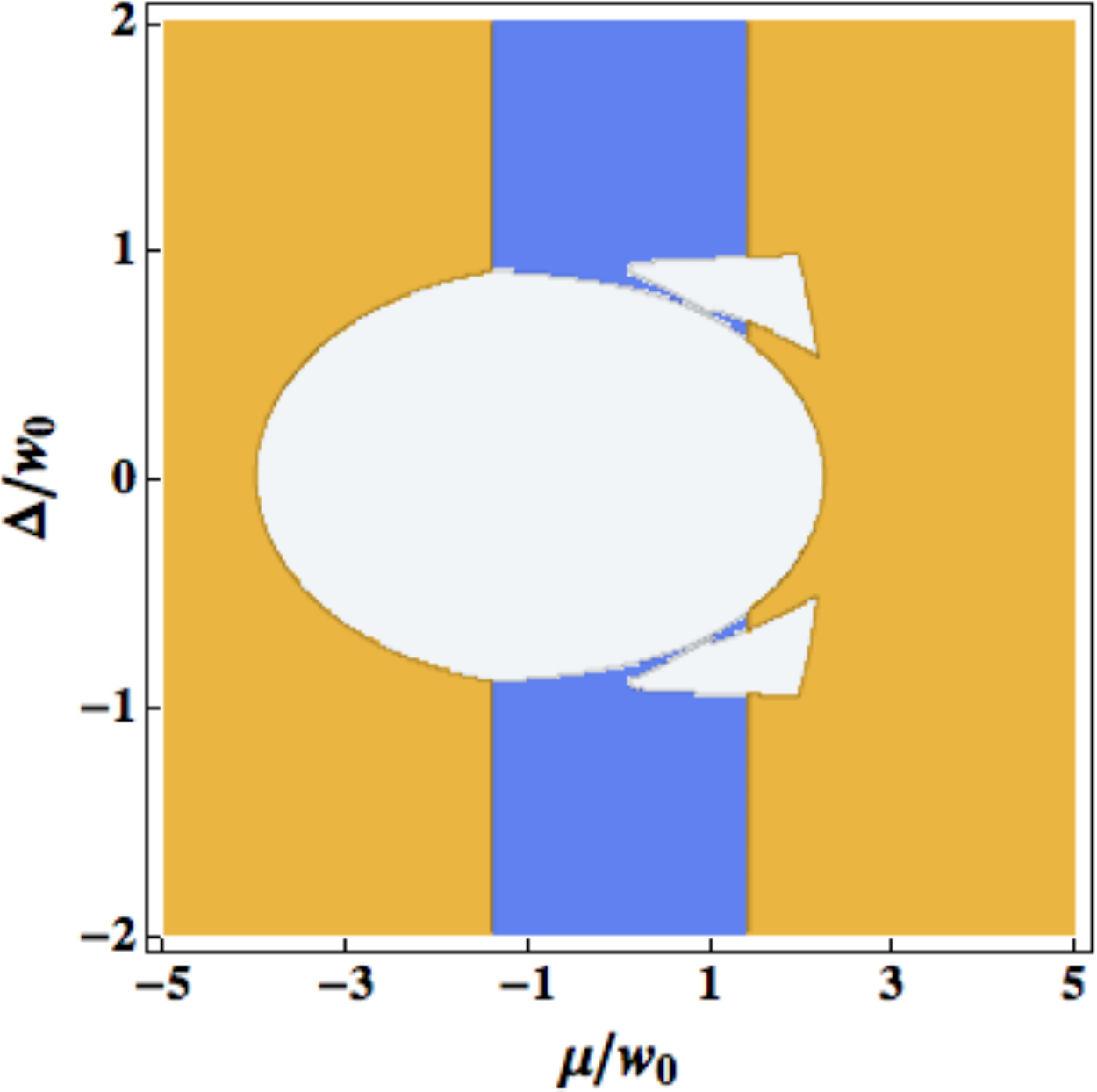}
                \label{fig6c2pi4}
                \includegraphics[width=0.4\columnwidth]{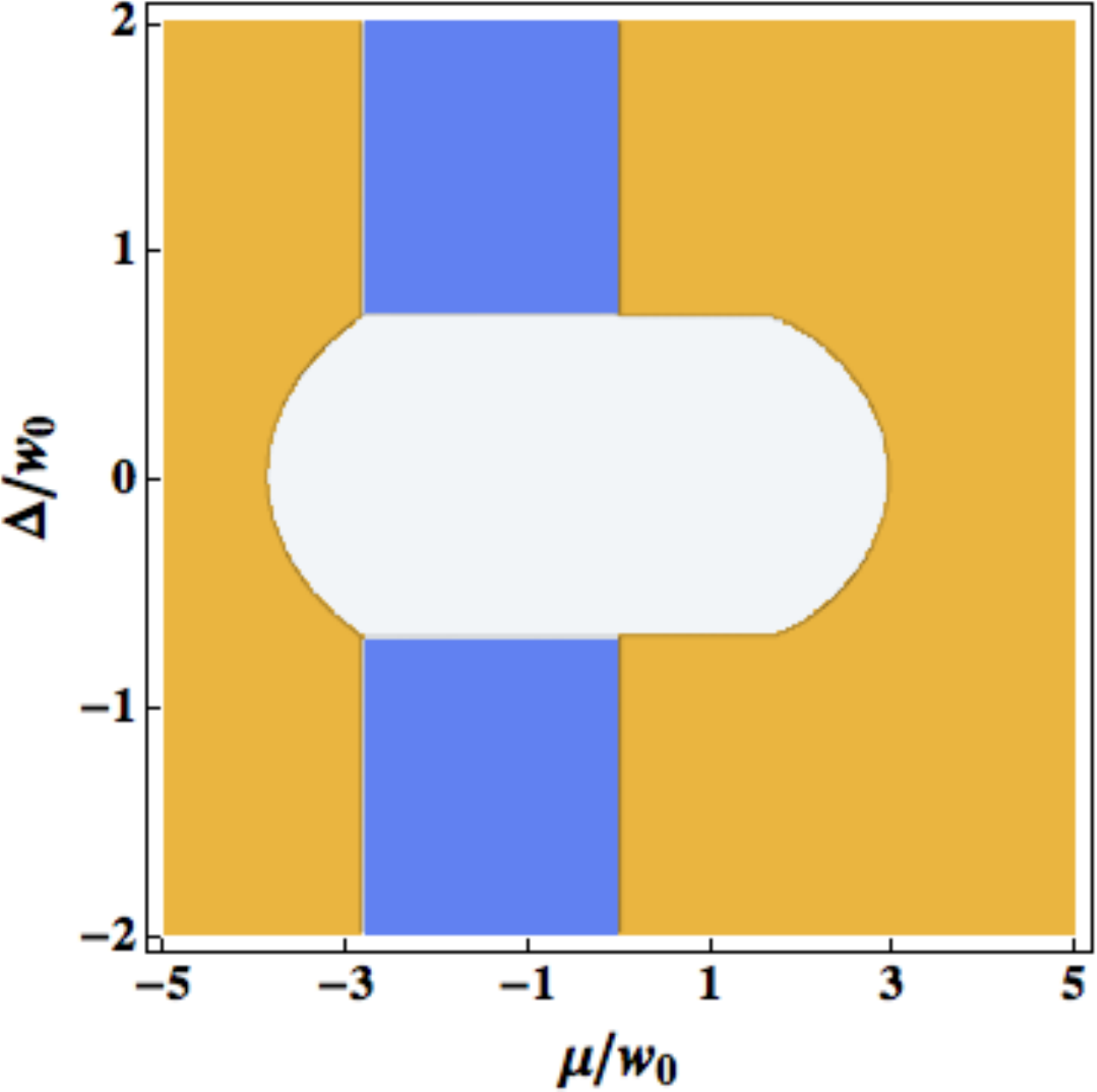}
                \label{fig6d2pi4}
   		\includegraphics[width=0.4\columnwidth]{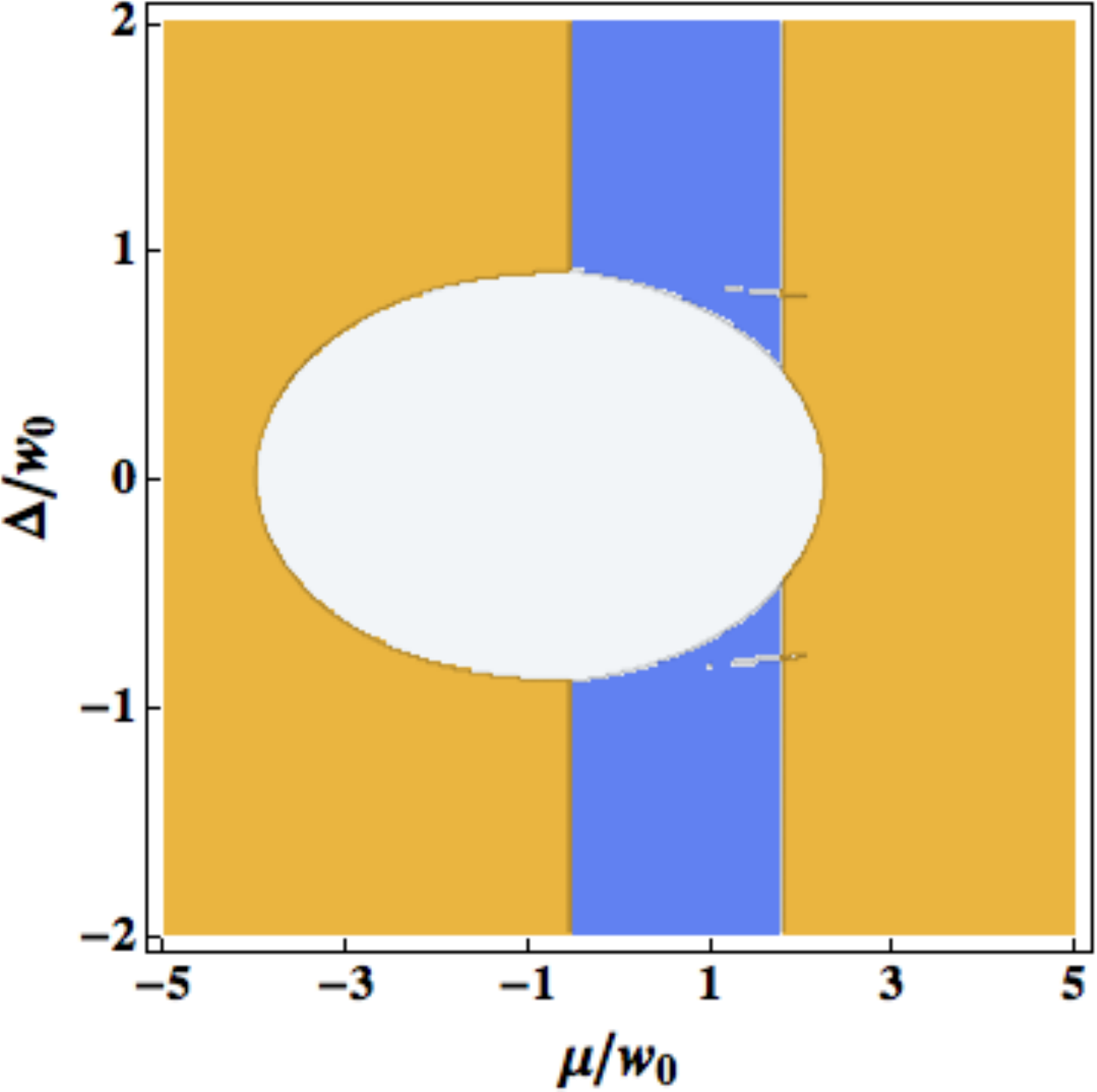}
   		\label{fig:BTRl3pisu10}
   		\includegraphics[width=0.4\columnwidth]{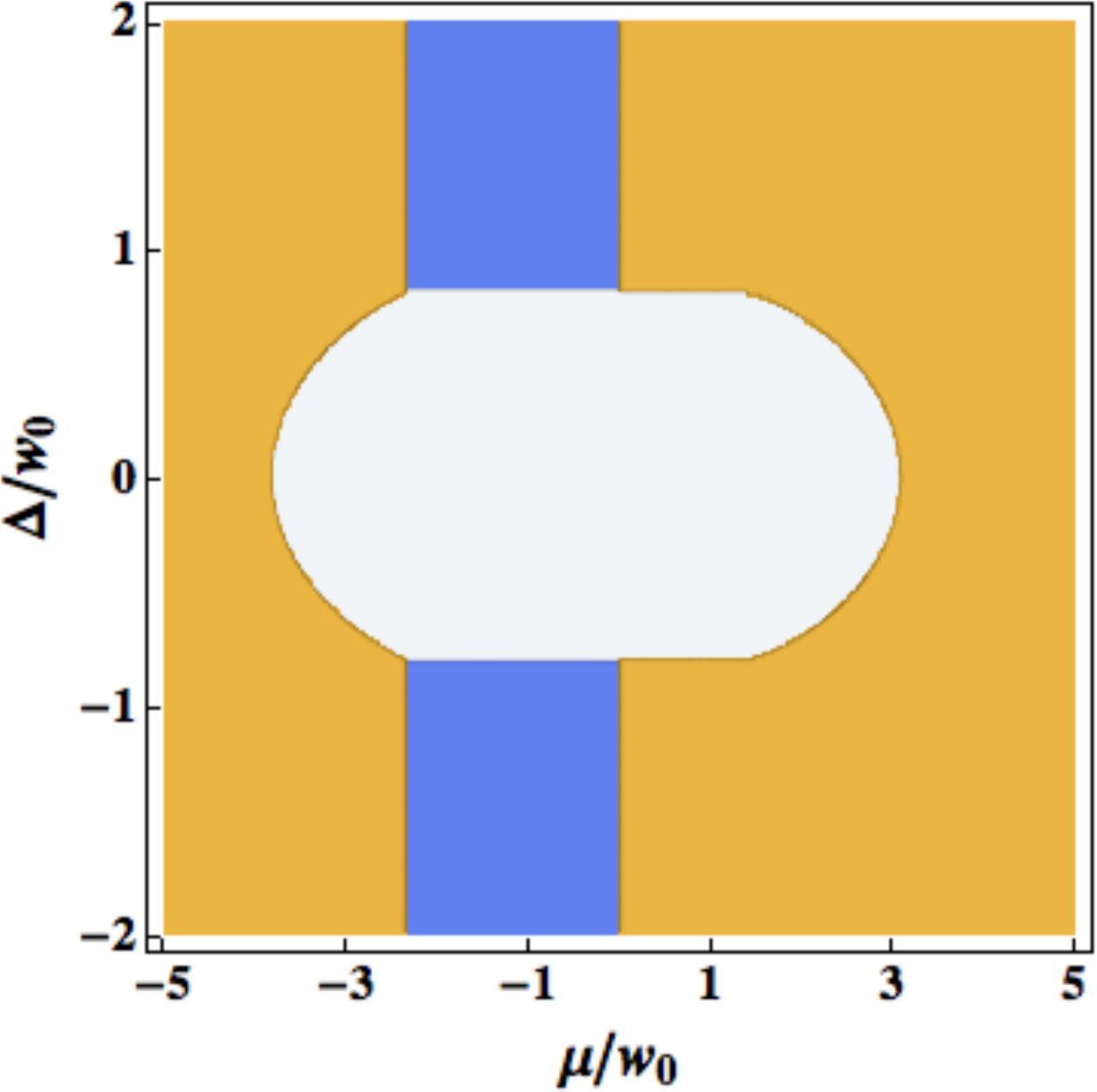}

                \label{fig:BTR3pisu10}
                \includegraphics[width=0.4\columnwidth]{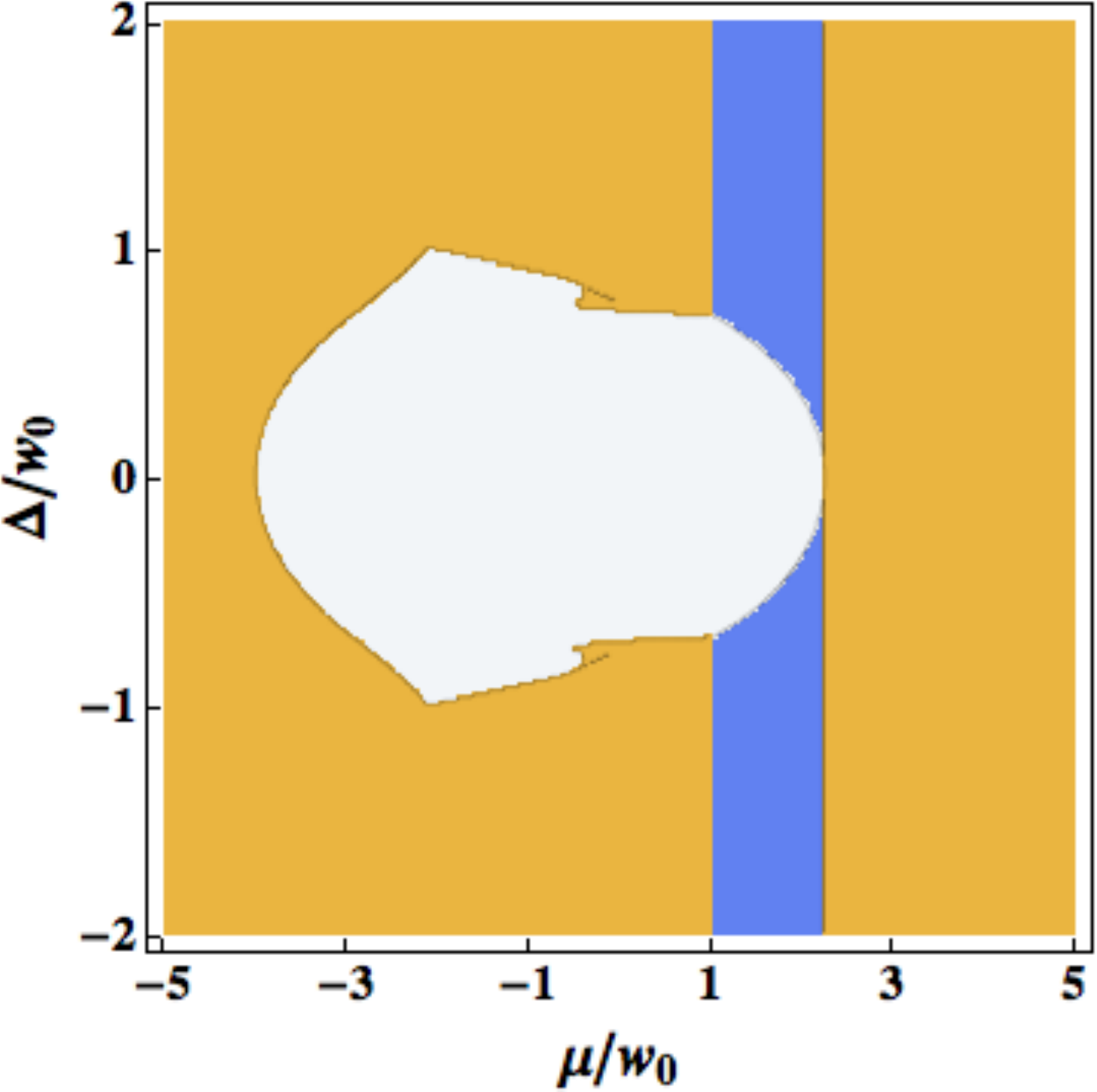}
                \label{fig6e2}
                \includegraphics[width=0.4\columnwidth]{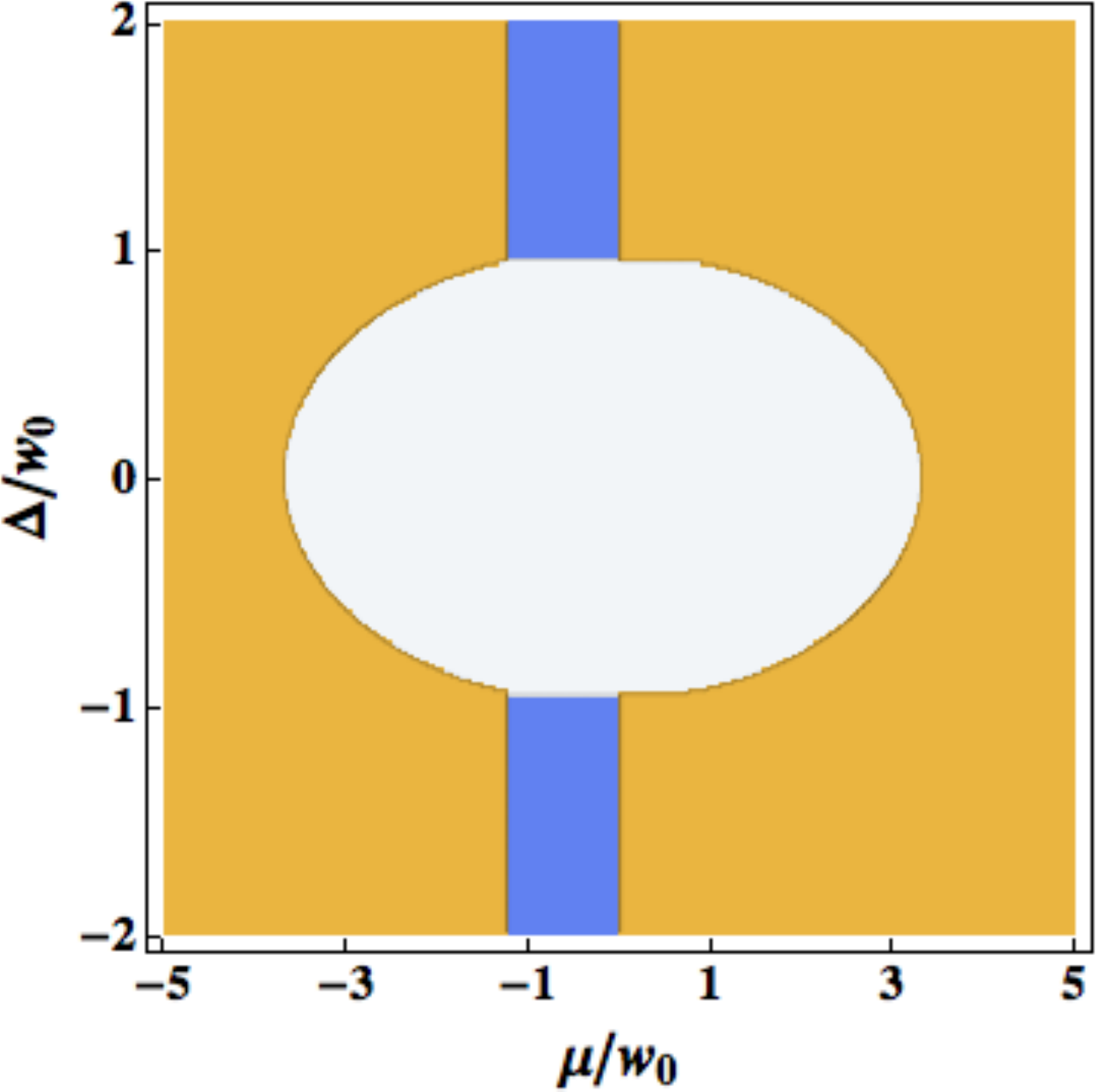}
   		\label{fig6f2}
   		\includegraphics[width=0.4\columnwidth]{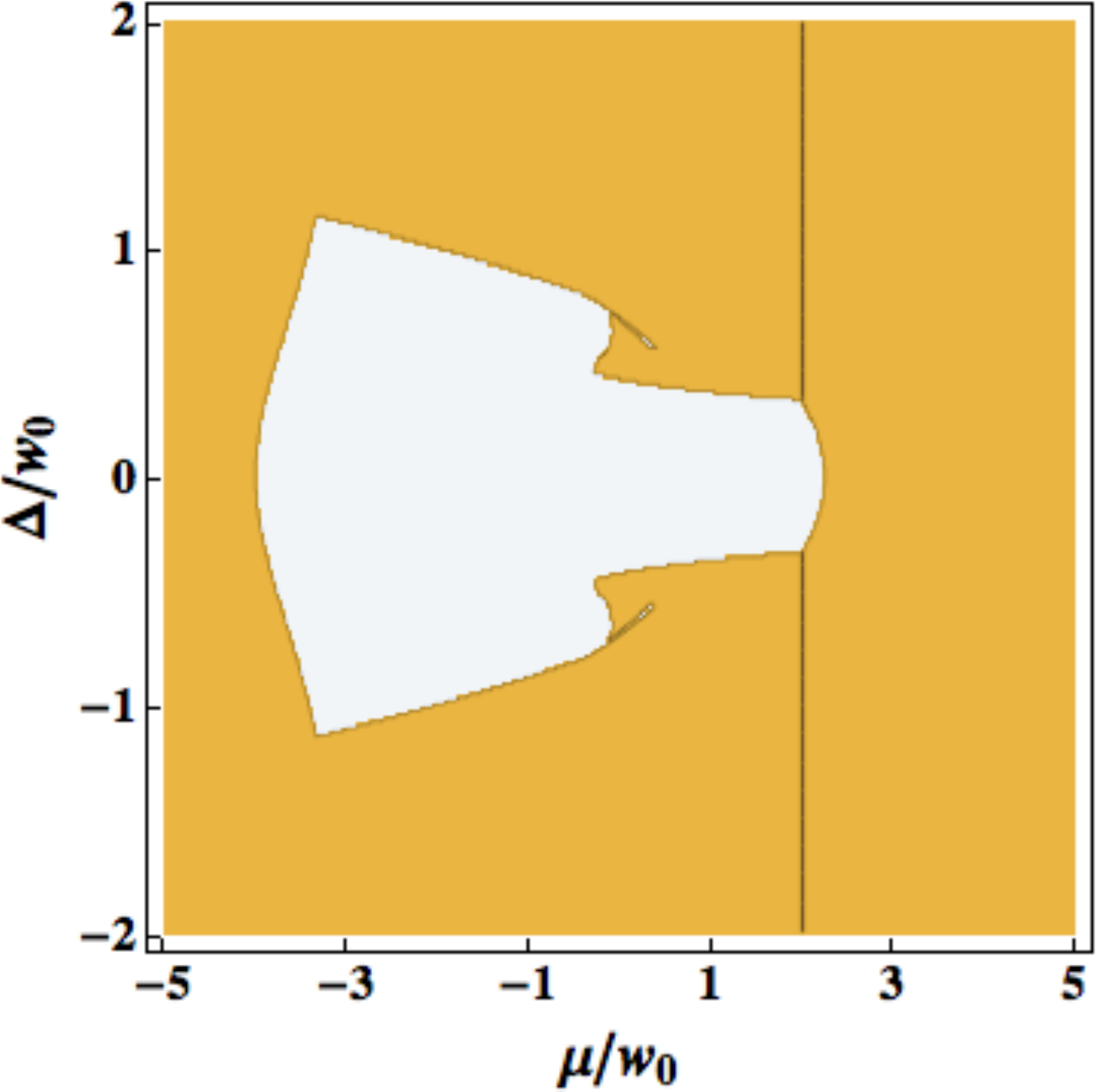}
   	\label{fig:BTRl5pisu10}
   		\includegraphics[width=0.4\columnwidth]{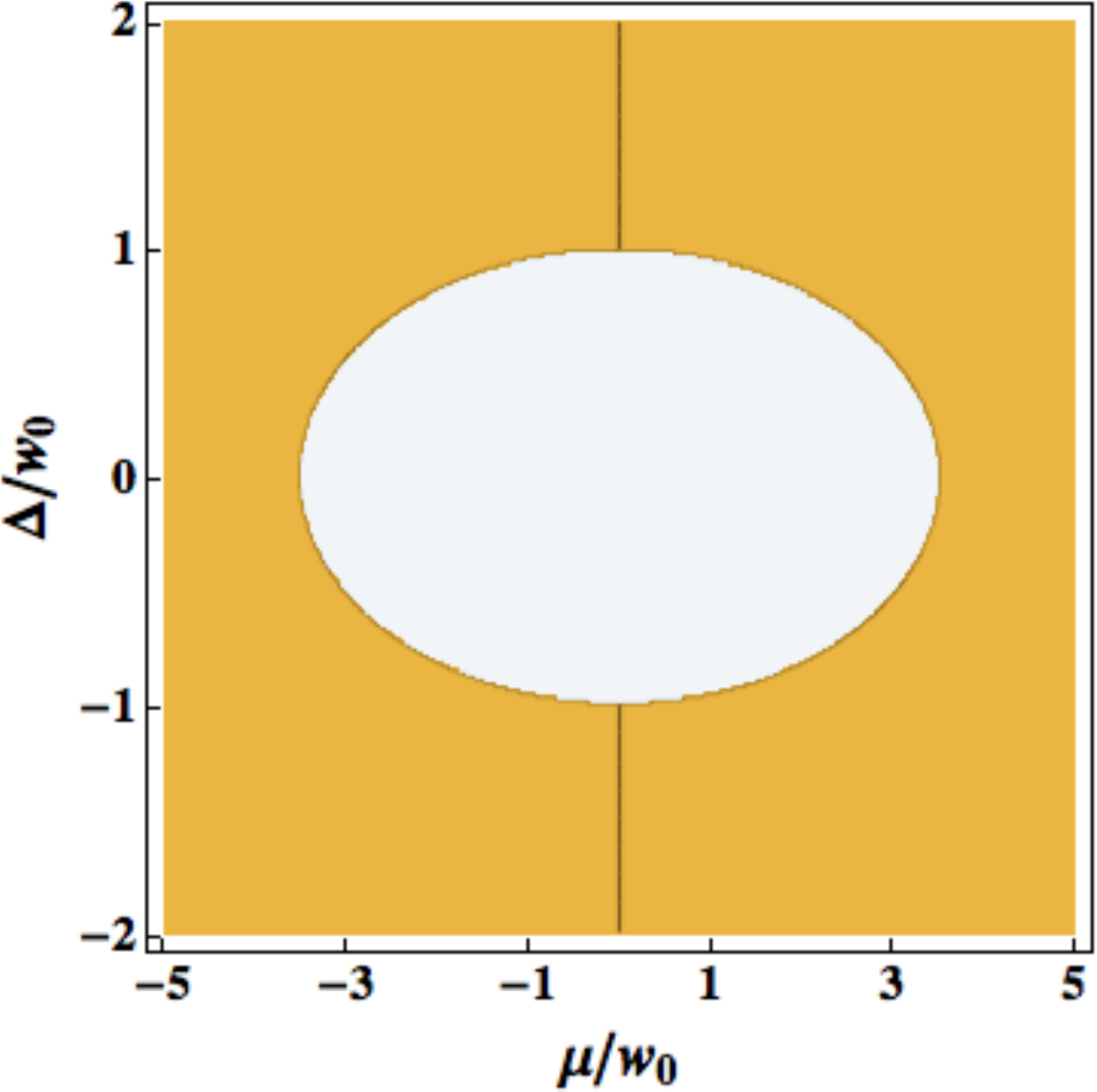}
   		\label{fig:BTR5pisu10} 
                \caption{Phase diagram for $\alpha=\beta=0$ for $r=2$ neighbor hopping and pairing, in the absence of time reversal symmetry. 
We consider two forms for the time reversal breaking phase $\varphi_\ell$: 
{\it i)} $\varphi_\ell=\varphi_0\ell$ (left), {\it ii)}
$\varphi_\ell=\varphi_0$ (right), for different $\varphi_0$, 
(from top to bottom) $\varphi_0=\pi/10, \pi/5, \pi/4, 3\pi/10, 2\pi/5, \pi/2$. The blue and yellow regions are respectively the topological and the trivial phases, while at their boundary and in the white critical regions the system is gapless.}
   	\label{fig:BTRr2}
   \end{figure}

 Including broken time reversal effects implies coupling between odd as well as even index Majorana operators so that 
only topological phase with an odd ${\sf w}$ will survive because of 
particle-hole symmetry of the spectrum \cite{DeGottardi}.
The time reversal symmetry is broken by the phase $\varphi_\ell$ in Eq.~(\ref{kspace}). 
We are going to study the effects of two different forms of this parameter: \emph{i)} $\varphi_\ell=\varphi_0$, $\forall \ell=1,\dots,r$ and \emph{ii)} $\varphi_\ell=\varphi_0\ell$ with 
$\ell=1,\dots,r$. 
Here we use the $\mathbb{Z}_2$ topological invariant 
$\upsilon=\sgn\left(h_z(0)h_z(\pi)\right)$ to detect the topological phase, 
where $h_z$ is the term multplying $\sigma_z$ in Eq.~\eqref{kspace}. 
It has two possible values, $\upsilon=-1$ if the phase is topological and $\upsilon=1$ if it is trivial.
The topological phase is permitted quite in general only if the condition
\begin{equation}
\label{top_condition}
-2\sum_{\ell=1}^rw_\ell\cos\varphi_\ell < \mu < -2\sum_{\ell=1}^r(-1)^\ell w_\ell\cos\varphi_\ell
\end{equation}
is fulfilled, which, in our case, becomes
\begin{equation}
\label{btr_top}
-2\sum_{\ell=1}^r\frac{\cos\varphi_\ell}{\ell^\alpha} < \frac{\mu}{w_0} <
-2\sum_{\ell=1}^r(-1)^\ell\frac{\cos\varphi_\ell}{\ell^\alpha}\,.
\end{equation}
For $\alpha=\beta=0$ and $\varphi_\ell=\varphi_0$ (if the phase does not depend on $\ell$, as in the plots on the r.h.s. of Fig.~\ref{fig:BTRr2}), then 
Eq.~(\ref{btr_top}) is given by $2 r\cos\varphi_0 < \mu/w_0 < 2\cos\varphi_0 \big(1+(-1)^r\big)/2$, while for $\alpha=\beta=0$ and 
$\varphi_\ell=\varphi_0\,\ell$ (as in all the plots 
shown for different $\varphi_0$ on the l.h.s. of Fig.~\ref{fig:BTRr2}) then Eq.~(\ref{btr_top}) 
becomes $-2\cos\big[(r+1)\varphi_0/2\big]\sin(r\varphi_0/2)/\sin(\varphi_0/2)< \mu/w_0 < \left\{1+\cos\big[(r\hspace{-0.05cm}+\hspace{-0.05cm}1)
(\varphi_0\hspace{-0.05cm}+\hspace{-0.05cm}\pi)\big]
+\tan(\frac{\varphi_0}{2})\sin\big[(r\hspace{-0.05cm}+\hspace{-0.05cm}1)
(\varphi_0\hspace{-0.05cm}+\hspace{-0.05cm}\pi)\big]\right\}$. 
The parametric regime where  topological states is permitted 
 shrinks as $\varphi_0$ increases, and disappears completely for $\varphi_0=\pi/2$, as one can easily check by putting this value on the expressions above.  
Let us now write the bulk energy-momentum dispersion 
\begin{eqnarray}
&&\hspace{-0.45cm}E_k^{\pm}=\sum_{\ell=1}^r w_\ell\sin\varphi_\ell\sin(k\ell)\\
&&\nonumber\pm\sqrt{\Big[\sum_{\ell=1}^r w_\ell
\cos\varphi_\ell\cos(k\ell)+\frac{\mu}{2}\Big]^2+\Big[\sum_{\ell=1}^r
\Delta_\ell \sin(k\ell)\Big]^2}
\end{eqnarray}
We find extended two-dimensional critical regions of parameters where $E_k^{\pm}$
 vanishes for some values of $k$ 
(white regions in Fig.~\ref{fig:BTRr2}), as in 
the case of short-range Kitaev chain with broken time reversal symmetry 
\cite{DeGottardi}. In particular, for $\varphi_\ell=\varphi_0\ell$, we can 
have also disconnected critical regions. These quantum critical points 
can be defined as the values of $\mu/w_0$ and $\Delta/w_0$ such that the following quantity 
\begin{equation}
\eta\equiv\min_{\{k\}}(E_k^+E_k^-)\max_{\{k\}}(E_k^+E_k^-)
\end{equation}
is negative. In other words the critical regions can be defined by 
$\sgn(\eta)=-1$, while the system is gapped if $\sgn(\eta)=1$.     

 \section{Infinite number of neighbors: topological phase diagrams}
Here we will consider an infinite number of 
neighbors, $r\rightarrow\infty$, first 
only in the pairing term, as done in Refs.~\cite{Lepori2} and \cite{delgado}, 
extending the analysis also for broken time reversal symmetry, 
and, afterwards, considering long range in both hopping and pairing terms.  
We will consider open boundary conditions, and perform exact diagonalization 
of the Hamiltonian Eq.~(\ref{hnostra}), with $r,\,L\rightarrow \infty$ (numerically $L\gg 1$). 
\change{Looking at the lowest energy levels}, 
we find that, together with massless Majorana modes, we obtain massive edge states, as predicted for long-range pairing \cite{Lepori2}, separated by a finite gap, therefore also called topological Dirac fermions \cite{delgado}. 

 \subsection{Long-range pairing, with and without TR symmetry}
In this section we will reconsider the Kitaev chain with long-range pairing, 
$w_\ell=w_0\delta_{\ell,1}$ and $\Delta_\ell=\Delta d_\ell^{-\alpha}$, 
in the presence of time reversal symmetry ($\varphi_\ell=\varphi_0= 0$), or in its absence ($\varphi_0\neq 0$)

  \subsubsection{With time reversal symmetry}
\begin{figure}[ht]
 	\centering 
 	\includegraphics[width=0.7\columnwidth]{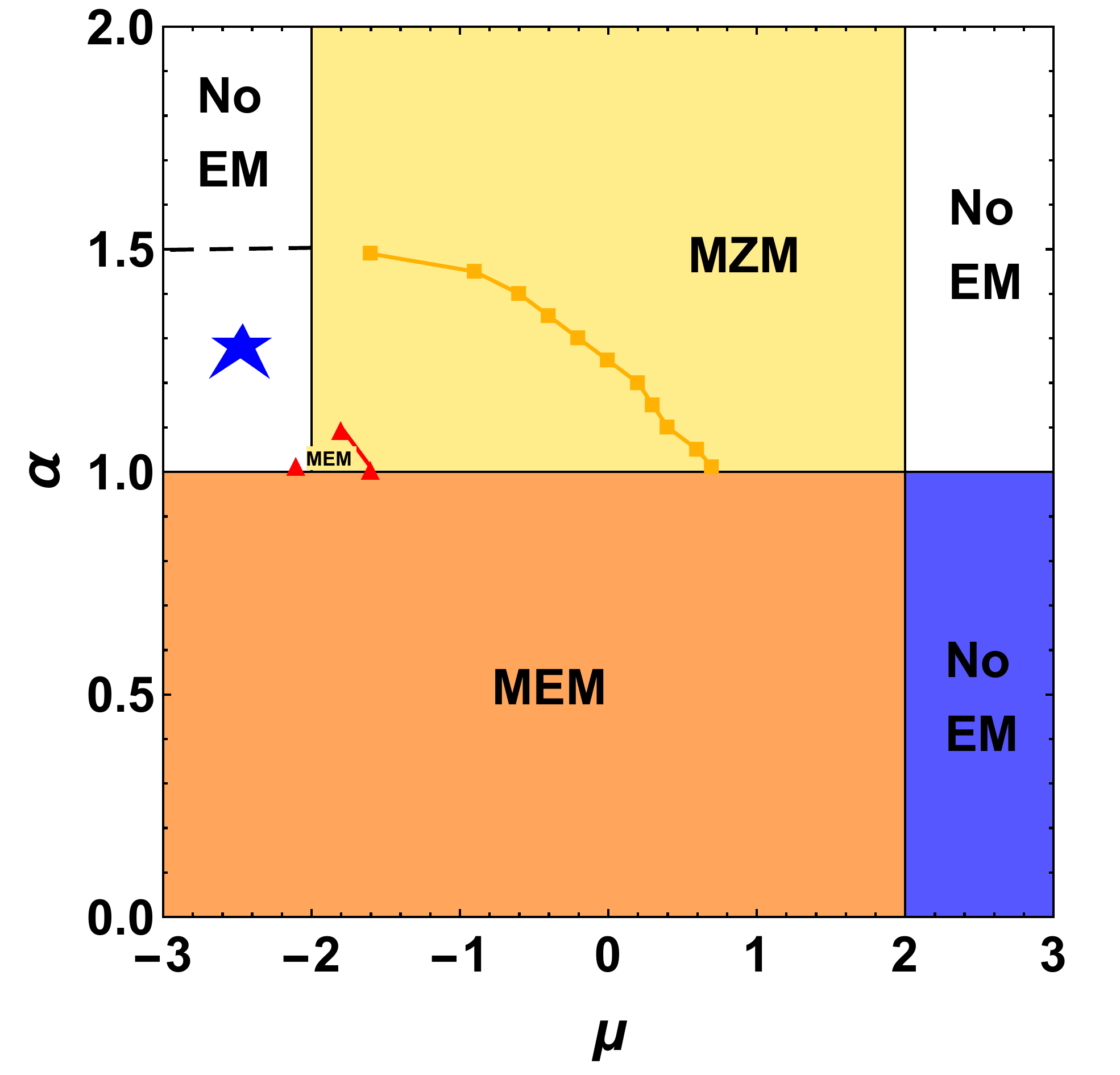}
\caption{Phase diagram for the long-range pairing model, with time reversal symmetry. We choose $\Delta=2w_0=2$. 
In the red region defined by $\mu< 2$ and $\alpha<1$, the winding number, using the definition in Eq.~(\ref{winding}) turns to be ${\sf w}=1/2$. In that regime the edge modes are massive (MEM), separated by a finite gap from the bulk 
modes. For $\mu>2$ and $\alpha<1$ there are no edge modes (no EM) and ${\sf w}=-1/2$. For $\alpha>1$, ${\sf w}=1$ for $-2 < \mu < 2$ and ${\sf w}=0$ 
otherwise.  
For $\mu\lesssim 1$ and $1\le \alpha\le 1.5$ one can find massive and 
massless edge modes. In the region with the star symbol the scaling analysis hardly converges up to $L\sim 10^{4}$,
thus we  can not give any information about the presence of edge modes. 
For $\alpha>1.5$ the system behaves like the standard Kitaev model.}
 	\label{fig:pairinglr}
 \end{figure}

\begin{figure}
        \centering
   \includegraphics[width=0.85\columnwidth]{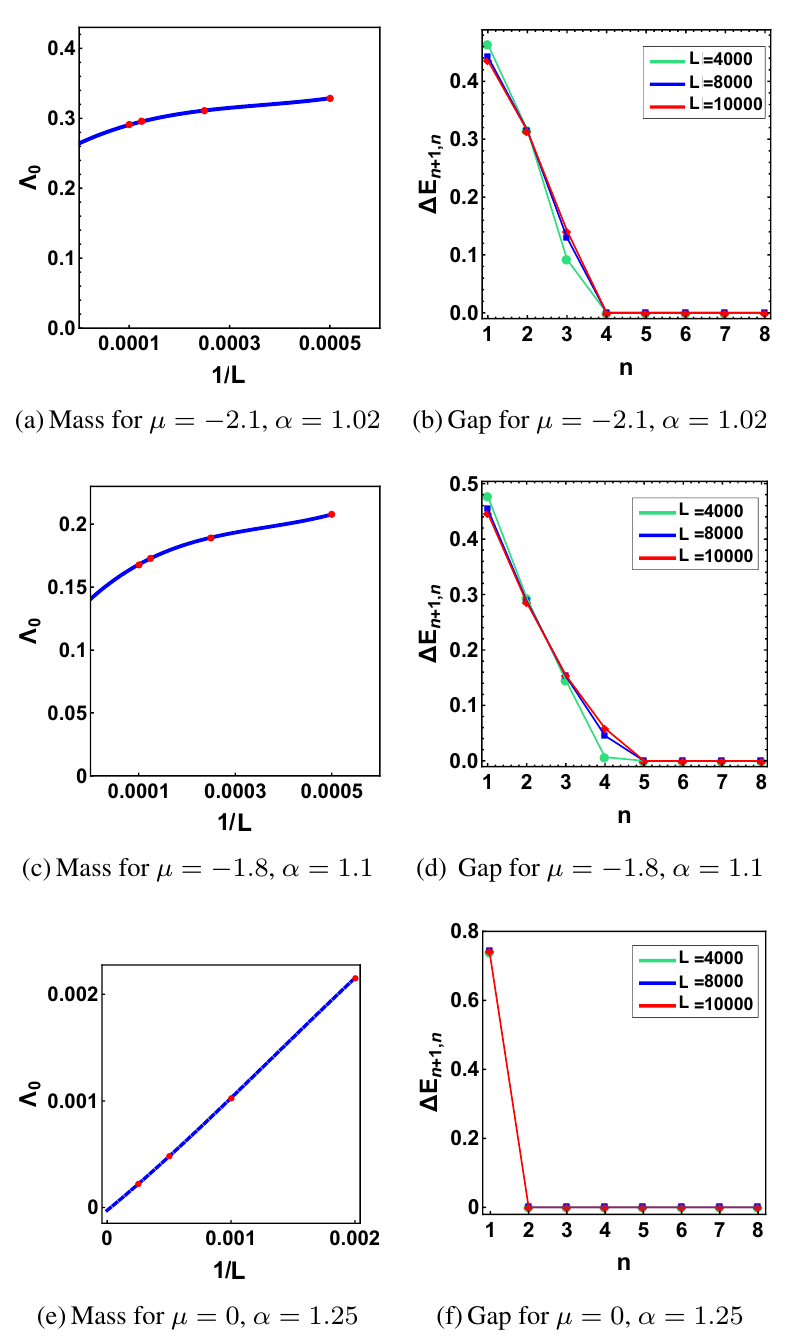}
%
 	\caption{Finite-size scaling for the mass ($\Lambda_0=E_1$, the first energy level) and the energy gaps ($\Delta E_{n+1,n}=E_{n+1}-E_n$), 
for three different points in the diagram in Fig.~\ref{fig:pairinglr}, 
the first two have finite masses and finite gaps, corresponding 
to massive edge modes (MEM), while in the last case the mass goes to zero in the thermodynamic limit while the gap stays finite, corresponding to a zero mode (MZM). We notice that in the first two cases not only the first level is separated by an energy gap but also the lowest three levels are gapped. 
An analog result is found also for broken time reversal symmetry case.}
 	\label{fig:MEMscaling}
 \end{figure}

Let us distinguish three regimes of values for $\alpha$, 
which correspond to the regimes where i) ${\cal H}$ and its derivates are not defined in $k=0$, ii) only the derivates of ${\cal H}$ are not defined in $k=0$, and iii) both the Hamiltonian and its derivates are defined over all the Brilloiun zone. 
For $\alpha <1$ due to the discontinuity in $k=0$ of ${\cal H}(k)$, 
the winding number ${\sf w}$ as defined in Eq.~(\ref{winding}), 
takes semi-integer values ${\sf w}=\pm \frac{1}{2}$ (see more comments about such winding numbers 
in Sec.~\ref{subsec_lhptr}). 
For $\mu< 2$ and $\alpha<1$ (here we choose $w_0=1$), the system has massive Dirac 
fermions at the edges, called massive edge modes (MEM),   
topologically protected by fermion parity and by a finite gap from bulk 
excitations, as shown in Ref.~\cite{delgado}.
The topological nature of the non-local massive Dirac fermions purely comes 
from the long-range deformation of the original Kitaev model. 
For $1<\alpha<1.5$ there is a cohexistence of massless edge modes 
(Majorana zero modes, MZM) and MEM, 
as already pointed out in Ref.~\cite{delgado}.
In this crossover region, we would like to give a more precise collocation of 
such massive and massless edge modes in the parameter space. 
We note that the presence of MEM also in the regions with ${\sf w}=0$ and 
${\sf w}=1$, for points closed to $(\mu,\alpha)=(-2,1)$, as shown by the 
red-triangular points in Fig.~\ref{fig:pairinglr}, are consistent with the numerical scaling analysis 
for the mass (ground state energy $\Lambda_0\equiv E_1$) and the first gap ($\Delta E_{21}=E_2-E_1$) (see Fig.~\ref{fig:MEMscaling} for some examples). 
Quite interestingly, we find that in many cases there are finite gaps in the thermodynamic limit also between a few of energy levels, 
see Figs.~\ref{fig:MEMscaling} (b), (d). 
We also checked other 
points (dark-yellow square points in Fig.~\ref{fig:MEMscaling}) where the mass is quite small, compatible with MZM. 
In the region signed by a blue star in Fig.~\ref{fig:MEMscaling} the scaling 
analysis is more difficult and not conclusive, therefore we cannot say anything about the edge modes there. 
However, this finding suggests that the winding number 
seems to be not enough to detect, for infinite-range model, 
the appearance of MZM, since at ${\sf w}=1$ one can find also MEM. 
For $\mu>2$ and $\alpha<1$ (blue region in Fig.~\ref{fig:MEMscaling}) we 
get ${\sf w}=-1/2$ which \change{seems} therefore topologically not equivalent to the 
trivial phase with ${\sf w}=0$, for $\alpha>1$, 
\change{(see comments on this in Sec.~\ref{subsec_lhptr})} 
even if in both cases there 
are no edge modes (no EM). For $\alpha>3/2$ the system behaves like the 
short-range Kitaev model, with MZM for $-2 < \mu < 2$. 
  \subsubsection{With broken time reversal symmetry}
\begin{figure}
 	\centering 
 	\includegraphics[width=0.7\columnwidth]{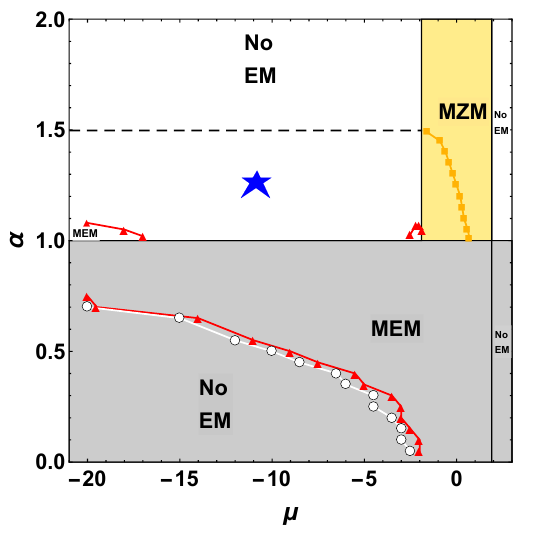}
 	\caption{Phase diagram for long-range paring model with broken time reversal symmetry ($\varphi_0=\pi/10$). Below $\alpha=1$ there is a critical line dividing massive edge mode (MEM) phase from the phase where there are no edge modes (No EM) depicted by red trangles (MEM) and white circles (No EM). The dark-yellow points (and the yellow region) denote MZM while 
In the region with the star symbol the scaling analysis hardly converges, thus we  can not give any information about the presence of edge modes.}
 	\label{fig:pairinglrbrokenTR}
 \end{figure}

In the broken time reversal symmetry case, the topological 
invariant $\upsilon$, Eq.~(\ref{upsilon}), is $\upsilon=-1$ for $-2w_0\cos\varphi_0 < \mu < 2w_0\cos\varphi_0$ and $\upsilon=1$ otherwise. 
For $\alpha>1.5$ it coincides with the presence or the absence of MZM, as shown in Fig.~\ref{fig:pairinglrbrokenTR}. For $1\le \alpha\le 1.5$ as in the previous case with time reversal symmetry, there is a coexistence of MZM and MEM 
together with regions where edge modes are absent. 
Finally in the regime with $\alpha<1$ the MZM seems to disappear and there is
a wide region where MEM are present. However the larger the negative $\mu$ 
the greater is the $\alpha$ below which the edge modes disappear. 
We numerically discover a critical line, for $\alpha<1$, in the phase diagram, 
shown in Fig.~\ref{fig:pairinglrbrokenTR}, separating the region with 
massive edge modes from the region where there are no edge modes. 

 \subsection{Long-range hopping and pairing, with and without TR symmetry}
In this section we will consider the Kitaev chain with same long range in the hopping, 
$w_\ell=w_0d_\ell^{-\alpha}$ and in the pairing $\Delta_\ell=\Delta d_\ell^{-\alpha}$ terms, in the presence of time reversal symmetry ($\varphi_\ell=0$), or in its absence ($\varphi_\ell\neq 0$). 
Quite in general, the Hamiltonian can be written as it follows 
\begin{equation}
\label{Hlong}
{\cal H}(k)=\begin{pmatrix}
h_0(k)+h_z(k)&-i h_y(k)\\
i h_y(k)&h_0(k)-h_z(k)
\end{pmatrix}
\end{equation}
where $\change{h_0(k)}=\sum_{\ell=1}^\infty w_\ell \sin\varphi_\ell \sin(k\ell)$, $\change{h_z(k)}=-{\mu}/{2}
-\sum_{\ell=1}^\infty w_\ell \cos\varphi_\ell \cos(k\ell)$ and $\change{h_y(k)}=\sum_{\ell=1}^\infty \Delta_\ell \sin(k\ell)$, which, taking the form $\varphi_\ell=\varphi_0+\varphi\, \ell$ for the phase, are
\begin{eqnarray}
\label{polyz}
&&\hspace{-0.2cm}h_0(k)\pm h_z(k)=\\ 
&&\nonumber  \mp \frac{1}{2}\left\{\mu+{w_0}\left[e^{-i\varphi_0}{\textrm{Li}}_\alpha(e^{-i(\varphi\pm k)})+e^{i\varphi_0}{\textrm{Li}}_\alpha (e^{i(\varphi\pm k)})\right]\right\}\\
&&\hspace{-0.2cm}h_y(k)\hspace{-0.1cm}=\hspace{-0.1cm}
i\frac{\Delta}{2}\left[{\textrm{Li}}_\alpha (e^{-ik})-{\textrm{Li}}_\alpha (e^{ik})\right]
\label{polyy}
\end{eqnarray}
where ${\textrm{Li}_\alpha (z)}$ is the polylogarithm of order $\alpha$ and argument $z$.  
 \subsubsection{With time reversal symmetry}
\label{subsec_lhptr}
For the long-range case the expression for the winding number ${\sf w}$ defined in Eq.~\eqref{winding}, 
because of the non-analycity of ${\cal H}(k)$, Eq.~(\ref{Hlong}), at $k=0$ for $\alpha<1$, gives, as in 
the long-range pairing case, as a result ${\sf w}=\pm\frac{1}{2}$.  
However we can draw the critical lines as the $r\rightarrow \infty$ limit of Eq.~(\ref{condhop}), 
requiring $h_z(0)=h_z(\pi)=0$, 
\begin{eqnarray}
&&\mu_{c1}=-2w_0\,\zeta(\alpha)\\
&&\mu_{c2}=2w_0\,\eta(\alpha)=2w_0(1-2^{1-\alpha})\zeta(\alpha)
\end{eqnarray}
defined by the Riemann zeta function and Dirichlet eta function. For $\alpha<1$, the winding number, from Eq.~(\ref{winding}) and using Eqs.~(\ref{polyz}), (\ref{polyy}), at $\varphi_\ell=0$, 
turns to be ${\sf w}=\frac{1}{2}$ for $\mu<\mu_{c1}$ and ${\sf w}=-\frac{1}{2}$ for $\mu>\mu_{c1}$. For $\alpha>1$ instead ${\sf w}=1$ for $\mu_{c1}<\mu<\mu_{c2}$ and ${\sf w}=0$ otherwise. 

At this point a comment about the winding number is in order. 
As in the case of long-range pairing, the winding number for $\alpha<1$ (long range) seems to be half 
an integer number because of the divergence of ${\cal H}(k)$, in terms of polylogharithmics, at $k=0$, and 
in particular because the function $h_y(k)$ does not close varying $k\in [0,2\pi)$. Actually we observe that 
\begin{eqnarray}
&&h_z(k)=h_z(2\pi-k)\,,\\
&&h_y(k)=-h_y(2\pi-k)\, .
\end{eqnarray} 
However if we perform the limit $k\rightarrow 0$ before $r\rightarrow \infty$, then $h_y(0)=h_y(\pi)=0$ therefore 
in this respect, by this regularization, the winding numbers come to be integer valued again, namely ${\sf w}=\pm\frac{1}{2}$ are replaced 
by \change{${\sf w}=1$, $0$}. 

\begin{figure}
 	\centering 
 	\includegraphics[width=0.7\columnwidth]{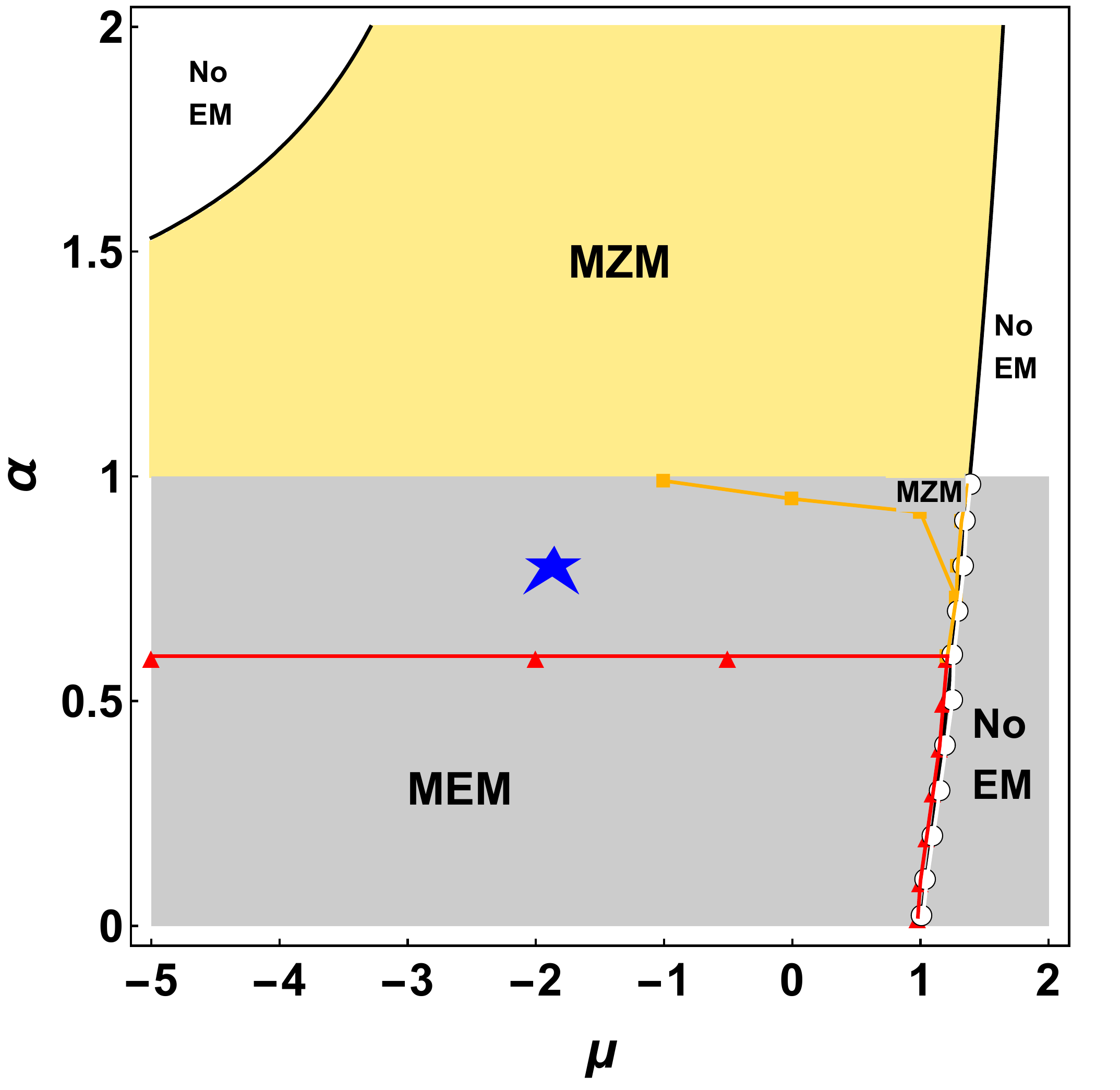}
\caption{Phase diagram for long-range hopping and pairing, with time reversal symmetry, for $\Delta=2w_0=2$. The yellow region is denoted by the topological invariant $\upsilon=-1$ and winding number ${\sf w}=1$, the white region by $\upsilon=1$ and ${\sf w}=0$ 
while in the grey region ($\alpha<1$) 
we can not define $\upsilon$ because of the divergence of $h_z(k)$ at $k=0$, while ${\sf w}=\frac{1}{2}$ for 
$\mu<2\eta(\alpha)$ and ${\sf w}=-\frac{1}{2}$ for $\mu>2\eta(\alpha)$ ($\mu=2\eta(\alpha)$ and $\mu=2\zeta(\alpha)$ the critical lines). 
For $\alpha>1$ MZM are allowed in the topological phase, while for $\alpha<1$ there are both MEM and MZM, at least for $\alpha$ slightly smaller than $1$ and 
$-1\lesssim \mu\lesssim 1$. For smaller values of $\alpha$, at least for $\alpha\lesssim 0.6$ and 
$\mu < 2\eta(\alpha)$, MEMs are dominant.}
 	\label{fig:longrangehpTR}
 \end{figure}

As far as the edge modes are concerned, for $\alpha\ge 1$ 
we get MZM for \change{$\mu_{c1} < \mu < \mu_{c2}$} and no edge modes otherwise. From finite-size scaling analysis on the 
energy spectrum we get 
edge modes with zero masses (MZM) also for $\alpha$ slightly below $1$, for $- w_0\lesssim \mu\lesssim 
2w_0\ln 2$. The regime with $\alpha\lesssim 0.6$ and $\mu<2w_0\eta(\alpha)$, instead, is characterized by massive edge modes (MEM), see Fig.~\ref{fig:longrangehpTR}.  
 \begin{figure}
 	\centering 
 	\includegraphics[width=0.7\columnwidth]{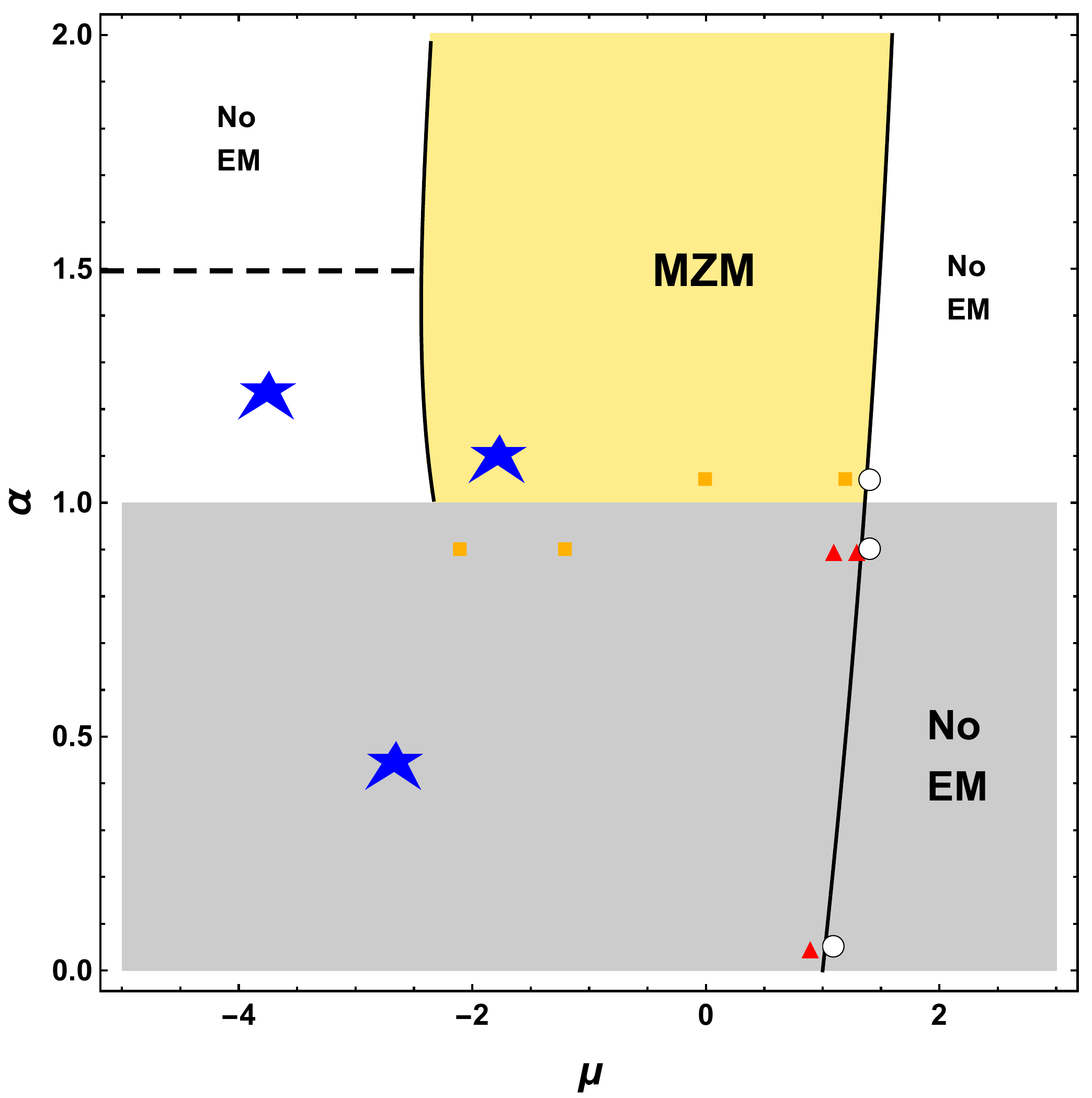}
\caption{Phase diagram for long-range hopping and pairing, with broken time reversal symmetry 
($\varphi_\ell=\ell \pi/10$). As in Fig.~\ref{fig:longrangehpTR}, we indicate the regions where MZM and MEM 
are present. At the dark-yellow square points correspond to MZM, the 
triangular red points to MEM, the white dots to no edge modes. 
The star symbols denote uncertainty about the edge modes.}
 	\label{fig:BTRlrphil}
 \end{figure}
 \subsubsection{With broken time reversal symmetry}
\begin{figure}
        \centering
\includegraphics[width=0.85\columnwidth]{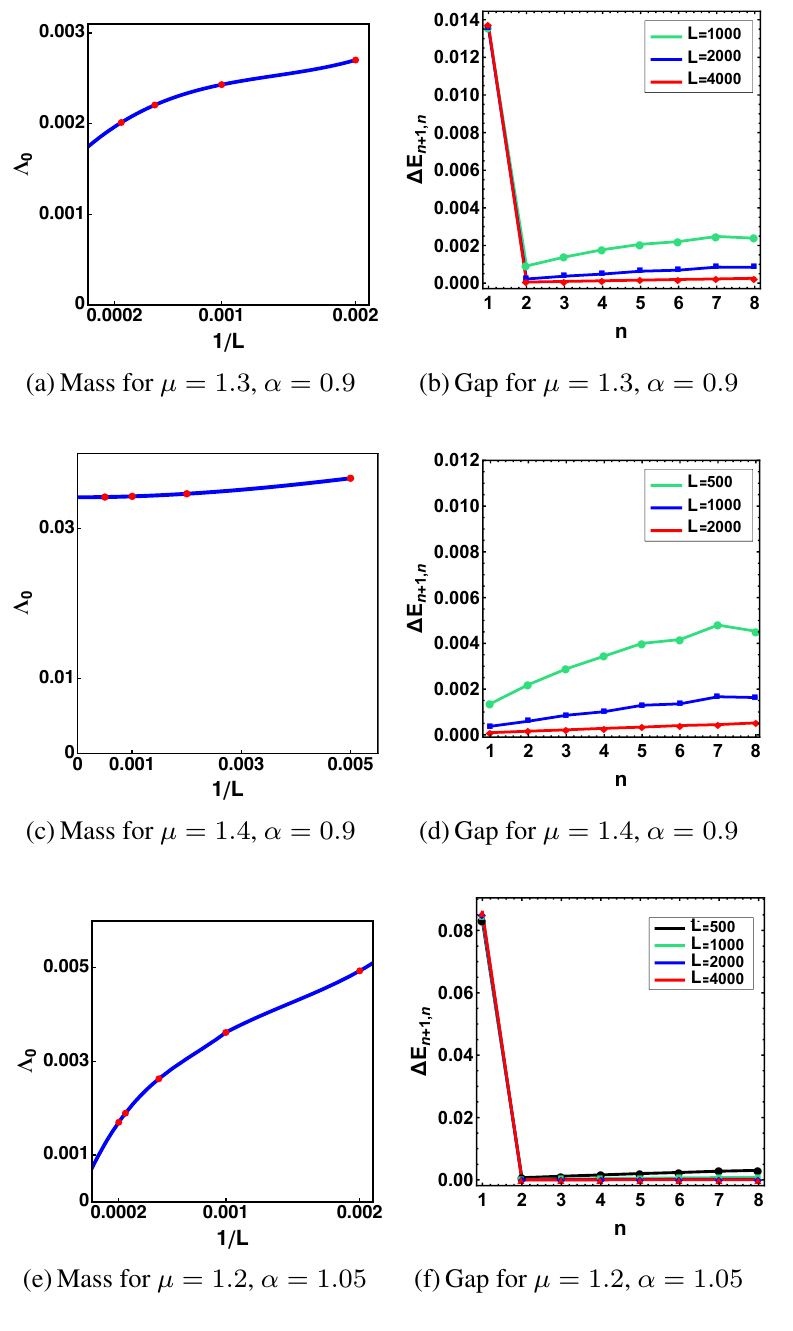}
\caption{Finite-size scaling for the mass ($\Lambda_0=E_1$, the first energy level) and the energy gaps 
($\Delta E_{n+1,n}=E_{n+1}-E_n$), for three different points in the diagram in Fig.~\ref{fig:BTRlrphil},
the first two have finite masses in the thermodynamic limit, (a), (c), 
but in the first case the gap is finite (b), corresponding to massive edge modes (MEM), 
while in the second case the gap goes to zero in the thermodynamic limit, corresponding to trivial phase. 
In the last case the mass seems to go to zero (up to $L=5000$) 
while the gap stays finite, a situation compatible with a zero mode (MZM).}
 	\label{fig:MEM}
 \end{figure}

We conclude this section considering the case of long-range hopping and pairing with broken time reversal 
symmetry, with $\varphi_\ell=\ell \varphi_0 \neq 0$. By this choice for the breaking symmetry phase, the critical 
lines are
\begin{eqnarray}
&&\mu_{c1}=-{w_0}\left[{\textrm{Li}}_\alpha(e^{-i\varphi_0})+{\textrm{Li}}_\alpha (e^{i\varphi_0})\right]\\
&&\mu_{c2}=-{w_0}\left[{\textrm{Li}}_\alpha (-e^{-i\varphi_0})+{\textrm{Li}}_\alpha (-e^{i\varphi_0})\right]
\end{eqnarray}  
so that the topological invariant is $\upsilon=-1$ for $\mu_{c1} < \mu < \mu_{c2}$ and $\upsilon=1$ otherwise. However we already experienced the fact that the topological invariants do not uniquely determine the 
presence of specific edge modes. By the analysis of the energy spectrum we get that for $\alpha>1$ the Majorana modes (MZM) are present in the topological phase denoted by $\upsilon=-1$. For $\alpha<1$ we can have MZM as well as MEM for $\mu<\mu_{c2}$, (see Fig.~\ref{fig:BTRlrphil}). Few examples of the finite-size scaling analyses are reported in Fig.~\ref{fig:MEM}.
 \section{Majorana wavefunctions}
In this section we calculate explicitly the wavefunctions of the zero energy modes for the 
extended Kitaev model with generic $r$-neighbor hopping and pairing terms, both in the thermodynamic limit,  
by means of a generalized transfer matrix approach, and 
for a finite length chain, generalizing the discussion done for the first-neighbor model in Ref.~\cite{Kitaev}. 
For this purpose it is more convenient to adopt the Majorana mode representation. 
Let us introduce and write the Majorana operators in terms of the fermionic operators 
$a_j$ and $a_j^\dagger$
 \begin{equation}
 c_{2j-1}=a_j+a^{\dagger}_j,\;\;\;\;\;
 c_{2j}
=i\big(a^{\dagger}_j-a_j\big)
 \end{equation}
which fulfill $\{c_i,c_j\}=2\delta_{i,j}\; \forall\,i,j=1,\dots,L$ 
and $c_j^{\dagger}=c_j$. 
The Hamiltonian Eq.~(\ref{hnostra}), therefore, can be rewritten in terms of the Majorana operators as follows
 \begin{eqnarray}
\label{ourMajo}
H &=&\frac{i}{2}\Big\{-\mu \sum_{j=1}^{L}
c_{2j-1}c_{2j}+\sum_{\ell=1}^{r}\change{\sum_{j=1}^{L-\ell}}\\
\nonumber&&\Big[-w_\ell\sin\varphi_\ell\,
\big(c_{2j-1}c_{2(j+\ell)-1}+c_{2j}c_{2(j+\ell)}\big)\\
\nonumber&&+(\Delta_\ell-w_\ell\cos\varphi_\ell)\,c_{2j-1}c_{2(j+\ell)}\\
\nonumber&&+(\Delta_\ell+w_\ell\cos\varphi_\ell)\,c_{2j}c_{2(j+\ell)-1}\Big]\Big\}
 \end{eqnarray}
We will now calculate the wavefunctions as solutions of the Bogoliubov-de Gennes equations (see Appendix \ref{app:bogoliubov}).
\subsection{Transfer matrix approach}
\begin{figure}[ht!]
 	\centering
 	\subfigure{
 		\includegraphics[width=.86\columnwidth]{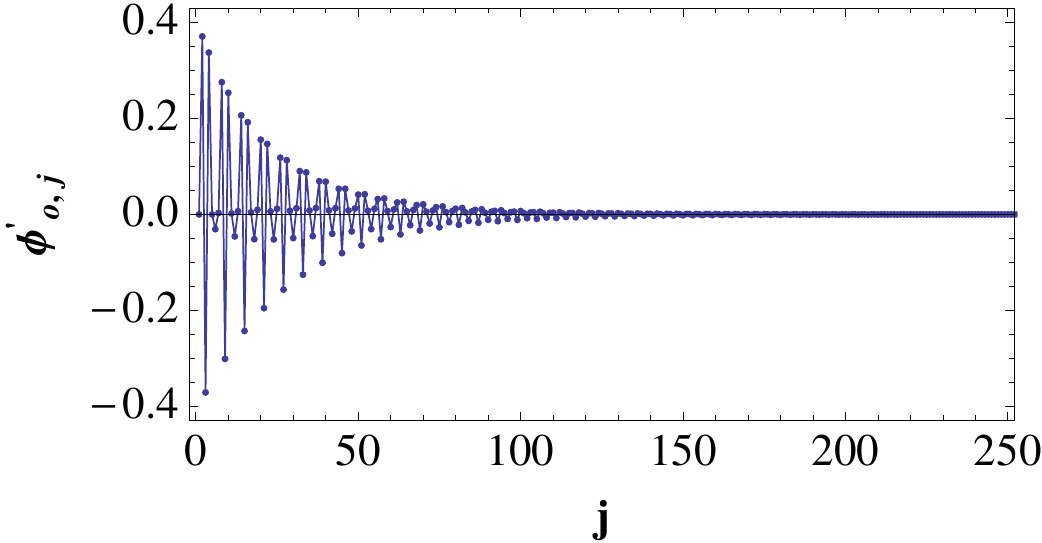}
 		\label{fig:fi1transfermatrix}}
 	\subfigure{
 		\includegraphics[width=.86\columnwidth]{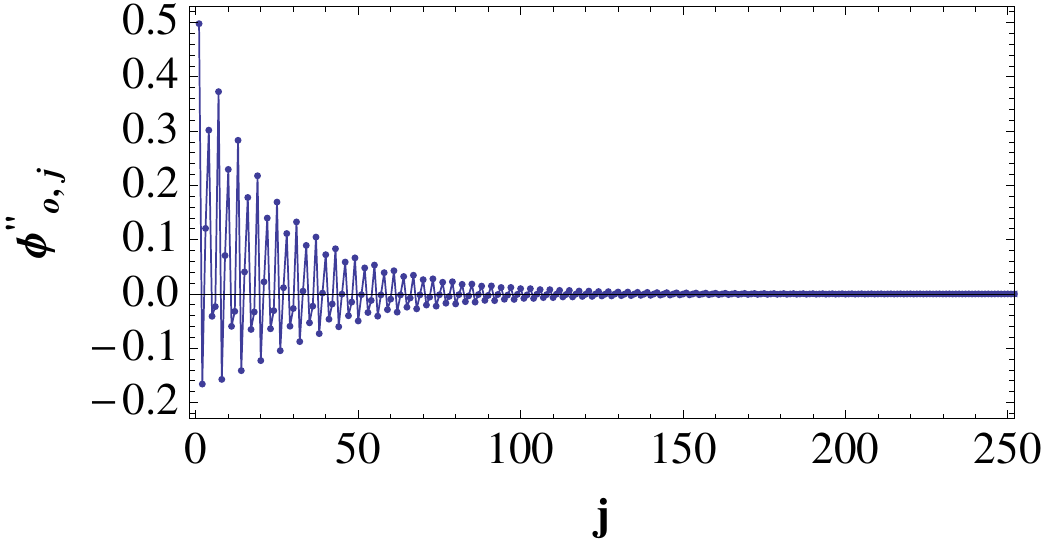}
 		\label{fig:fi2transfermatrix}}
\caption{Two ortonormalized wavefunctions localized at the same edge, 
namely two independent MZM per edge, 
in the exteded Kitaev chain for $r=2$ neighbor hopping and pairing, with $\Delta_1=\Delta_2=1$, $\mu=w_1=w_2=0.1$ (in terms of 
Eq.~(\ref{parameters}),   
$\alpha=\beta=0$, $\mu/\Delta=w_0/\Delta=0.1$), obtained by the transfer matrix approach.}
 	\label{fig:analiticonumerico} 
 \end{figure}

We will consider for simplicity the case with time reversal symmetry ($\varphi_\ell=0$), so that Eq.~(\ref{ourMajo}) reduces to 
\begin{eqnarray}
\label{HM}
H &=&\frac{i}{2}\Big\{-\mu \sum_{j=1}^{L}
c_{2j-1}c_{2j}+\sum_{\ell=1}^{r}\change{\sum_{j=1}^{L-\ell}}\\
\nonumber&&\Big[
(\Delta_\ell-w_\ell)\,c_{2j-1}c_{2(j+\ell)}
+(\Delta_\ell+w_\ell)\,c_{2j}c_{2(j+\ell)-1}\Big]\Big\}
\end{eqnarray}
Introducing the wavefunctions $\phi_{\epsilon_n,j}$ and $\psi_{\epsilon_n,j}$, 
related to the Bogoliubov coefficients which diagonalize $H$ ($\phi_{\epsilon_n,j}=u_{n,j}+v_{n,j}$, $\psi_{\epsilon_n,j}=u_{n,j}-v_{n,j}$), we get the following Bogoliubov equations (see Appendix \ref{app:bogoliubov} for details) 
\begin{eqnarray}
\label{phi}
\nonumber\sum_{\ell=1}^r\Big[
(\Delta_\ell-w_\ell)\phi_{\epsilon_n,j-\ell}
-(\Delta_\ell+w_\ell)\phi_{\epsilon_n,j+\ell}\Big]-\mu \phi_{\epsilon_n,j}\\
=\epsilon_n \psi_{\epsilon_n,j}\;\\
\nonumber\sum_{\ell=1}^r\Big[(\Delta_\ell-w_\ell)\psi_{\epsilon_n,j+\ell}
-(\Delta_\ell+w_\ell)\psi_{\epsilon_n,j-\ell}\Big]-\mu \psi_{\epsilon_n,j}\\
=\epsilon_n \phi_{\epsilon_n,j}\;
\label{psi}
\end{eqnarray}
\change{For the most general case one has to solve 
Eqs.~(\ref{W1})-(\ref{W4})}. 
Since we are interested in the zero energy states, we will consider the case 
where, for some ${n}$, we have $\epsilon_n=0$. \\
In this case the wavefunctions $\phi_{o,j}$ and $\psi_{o,j}$, related to zero energy, in Eqs.~(\ref{phi}), (\ref{psi}) decouple and we can solve the equations separately. 
Supposing $\Delta_r+w_r\neq 0$, Eqs.~(\ref{phi}), (\ref{psi}), for $\epsilon_n=0$, can be written in the following form 
 	\begin{eqnarray}
\label{transfer}
 \begin{pmatrix}
 \phi_{o,j+r}\\
 \vdots\\
 \phi_{o,j-r+1}
 \end{pmatrix}=T\begin{pmatrix}
 \phi_{o,j+r-1}\\
 \vdots\\
 \phi_{o,j-r}
 \end{pmatrix}\\
\nonumber\\
 \begin{pmatrix}
 \psi_{o,j-r}\\
 \vdots\\
 \psi_{o,j+r-1}
 \end{pmatrix}=T\begin{pmatrix}
 \psi_{o,j-r+1}\\
 \vdots\\
 \psi_{o,j+r}
 \end{pmatrix}
\label{transfer2}
 \end{eqnarray}
after introducing the following transfer matrix  
 \begin{equation}
\label{Tmatrix}
 T=\begin{pmatrix}
t_1&\dots&t_r&\dots&t_{2r}\\
 1&0&\hdotsfor{2}&0\\
 0&1&0&\dots&\vdots\\
 \vdots&\dots&\ddots&\ddots&0\\
 0&\hdotsfor{2}&1&0
 \end{pmatrix}
 \end{equation}
where
\begin{eqnarray}
&&t_{i}=-\frac{\Delta_{r-i}+w_{r-i}}{\Delta_r+w_r},\;\;\textrm{for}\;i=1,\dots,r-1\\
&&t_r=-\frac{\mu}{\Delta_r+w_r}\\
&&t_{i}=\frac{\Delta_{i-r}-w_{i-r}}{\Delta_r+w_r},\;\;\textrm{for}\;i=r+1,\dots,2r
\end{eqnarray}
It is straightforward to notice that $t_1={\textrm{Tr}}(T)=\sum_{i=1}^{2r}\lambda_i$, is the trace of $T$, and $t_{2r}=-\textrm{Det}(T)=-\Pi_{i=1}^{2r} \lambda_i$, the determinant, where $\lambda_i$ are the eigenvalues of $T$. \\ 
By this approach one can write the wavefunctions evaluated at some point from 
its value at another point by applying several times the transfer matrix, 
namely applying $T$ to some power, say $j$, related to the space distance 
between the two points. What is relevant is therefore $T^{j}$, or, more 
conveniently, its diagonal form, so that $T^j$ can be written as $T^j=
SD^jS^{-1}$ were $S$ diagonalizes $T$ ($D=S^{-1}TS$). The problem is reduced, 
therefore, to finding the eigenvalues ($D$) and the eigenstates ($S$) of $T$. 
In order to find the eigenvalues one has to write the charateristic polynomial 
$p_{2r}(\lambda)$ of the $2r\times 2r$ matrix in Eq.~(\ref{Tmatrix}), and find the solutions of $p_{2r}(\lambda)=0$. One can easly prove that the polynomial 
$p_{2r}(\lambda)=\textrm{Det}(T-\lambda \mathbb{1})$ is such that
\begin{equation}
p_{2r}(\lambda)=\lambda\, p_{2r-1}(\lambda)-t_{2r}
\end{equation}
and, therefore, by iteration, and making it equal to zero, one has to solve the following eigenvalue equation
\begin{equation}
\label{charpoly}
p_{2r}(\lambda)=\lambda^{2r}-\sum_{i=1}^{2r}t_i\,\lambda^{2r-i}=0
\end{equation}
in order to find the $2r$ eigenvalues of $T$, $\lambda_s$ with $s=1,\dots,2r$. 
One can easily check that the corresponding eigenfunctions are 
$(\lambda_s^{2r}, \lambda_s^{2r-1},\dots,\lambda_s,1)^t$, which are the 
columns composing the matrix $S$. As a result, 
generic solutions of Eqs.~\eqref{transfer}, \eqref{transfer2} 
can be written as the following linear combinations
\begin{eqnarray}
\label{clphi}
&&\phi_{o,j}=\displaystyle\sum_{s=1}^{2r}c^\phi_s\,\lambda_s^{j}\\
&&\psi_{o,j}=\displaystyle\sum_{s=1}^{2r}c^\psi_s\,\lambda_s^{-j}
\label{clpsi}
\end{eqnarray}
where 
the coefficients $c^\phi_s$ and $c^\psi_s$ are independent and $j=1,\dots,L$. 
A part of those coefficients can be fixed by imposing open 
boundary conditions. 
Let us call $n^{<}$ the number of $\lambda_s$ such that $|\lambda_s|<1$ and, 
analogously, $n^{>}$ the number of eigenvalues such that $|\lambda_s|>1$. 
Let us suppose the majority of $\lambda_s$ are smaller than $1$, namely if 
$n^< >n^>$, then the boundary conditions (open boundary condition and 
thermodynamic limit) require
\begin{eqnarray}
&&\phi_{o,0}=\phi_{o,-1}=\dots=\phi_{o,1-r}=0\\
&&\psi_{o,L+1}=\psi_{o,L+2}=\dots=\psi_{o,L+r}=0
\end{eqnarray}
In this way we can have functions localized at the edges, $\phi_{o,j}$ on the left (small $j$) and $\psi_{o,j}$ on the right (large $j$). 
The eigenvalues greater than one in modulus will be discarted ($c^\psi_s=c^\phi_s=0$) to satisfy normalization conditions.
The situation is reversed if, instead, $n^< < n^>$. In this case 
we have to impose the following boundary conditions 
\begin{eqnarray}
&&\phi_{o,L+1}=\phi_{o,L+2}=\dots=\phi_{o,L+r}=0\\
&&\psi_{o,0}=\psi_{o,-1}=\dots=\psi_{o,1-r}=0
\end{eqnarray}
verifying the existence of zero energy modes which will be localized at the edges, $\psi_{o,j}$ on the left and $\phi_{o,j}$ on the right.\\
In both the cases, the number of boundary conditions to impose for each 
wavefunction is $r$. The degrees of freedom to construct the wavefunctions in 
Eqs.~(\ref{clphi}),~(\ref{clpsi}) are, therefore, 
equal to $N=\big(\textrm{max}(n^<,n^>)-r\big)$. This means that we can 
construct 
$N$ different linear combinations Eqs.~(\ref{clphi}),~(\ref{clpsi}) at each edge, 
namely $N$ different 
zero-energy wavefunctions per edge. Since $\textrm{max}(n^<,n^>)\le 2r$, 
we can get at most $N=r$ Majorana states per edge, consistently with the
winding number analysis. A particular attention has to be paid for complex 
$\lambda_s$ since globally $\phi_{o,j}$ and $\psi_{o,j}$ have to be real 
valued by construction. This is also the reason of some oscillating behaviors 
of the wavefunctions, which come together with an overall exponential 
decay, since complex eigenvalues appear in pairs, therefore, $|\lambda_s|^je^{i\theta j}+|\lambda_s|^je^{-i\theta j}=2\,e^{j\ln|\lambda_s|}\cos(\theta j)$. \\
An example with $r=2$ is provided 
in Fig.~\ref{fig:analiticonumerico}, 
where two different $\phi_{o,j}$, two ortonormalized 
wavefunctions $\phi^{'}_{o,j}$, $\phi^{''}_{o,j}$, both localized at the same 
edge, has been found for a set of parameters for which the winding number is 
${\sf w}=2$. Analogously one can find two independent wavefunctions localized 
on the other edge of the chain. \\ 
Another simple example which shows the connection between this approach 
and the previous 
winding number analysis is the following. Let us consider for simplicity $r=2$ and $w_1=w_2=\Delta_1=\Delta_2=1$, 
or in terms of the parametrization in Eq.~(\ref{parameters}), 
$\alpha=\beta=0$ and $w_0=\Delta=1$. In this case Eq.~(\ref{charpoly}) is simply 
$\lambda^4+\lambda^3+\frac{\mu}{2}\lambda^2=0$ whose solutions are all $|\lambda_s|<1$ for $0< \mu<2$, consistently with Eq.~(\ref{wr}) for $r=2$. 
 
\change{Actually, the case $w_\ell=\pm \Delta_\ell$} 
has to be consider separately, as we will do in the next section, since in that case $T$ 
is singular and some of the eigenvalues are zero, requiring in some cases 
$\phi_{o,j}$ and $\psi_{o,j}$ to be Dirac delta functions, 
which are actually distributions, not really functions. 
The advantage in that case is that we can relax the condition of 
working in the thermodynamic limit, required for vanishing overlap between the 
wavefunctions localized at the two edges, which otherwise would hybridize 
spoiling the presence of unpaired Majorana modes.

 \subsection{Special cases: Majorana modes in a finite-lenght chain}
\label{finite}
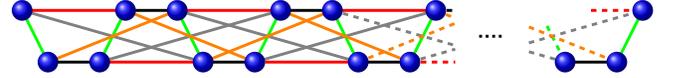
\begin{figure}
 	\resizebox{\columnwidth}{!}{
 	\begin{tikzpicture}
 	\draw[color=green, ultra thick] (0,0) -- (0.5,-1);
 	\draw[color=green, ultra thick] (1.5,-1) -- (2.0,0);
 	\draw[color=green, ultra thick] (3.0,0) -- (3.5,-1);
 	\draw[color=green, ultra thick] (4.5,-1) -- (5.0,0);
 	\draw[color=green, ultra thick] (6.0,0) -- (6.5,-1);
 	\draw[color=green, ultra thick] (7.5,-1) -- (8.0,0);
 	\draw[color=green, ultra thick, dashed] (10.15,-0.3) -- (10.5,-1);
 	\draw[color=green, ultra thick] (11.5,-1) -- (12,0);
 	\draw[ color=gray, ultra thick] (0,0) -- (3.5,-1);
 	\draw[ color=gray, ultra thick] (3,0) -- (6.5,-1);
 	\draw[ color=gray, ultra thick] (1.5,-1) -- (5,0);
 	\draw[color=gray, ultra thick] (4.5,-1) -- (8,0);
 	\draw[color=gray, ultra thick, dashed] (6,0) -- (8.37501,-0.678574);
 	\draw[color=gray, ultra thick, dashed] (7.5,-1) -- (8.37501,-0.75);
 	\draw[color=gray, ultra thick, dashed] (9.8,-0.8) -- (10.5,-1);
 	\draw[color=gray, ultra thick, dashed] (9.8,-0.628571) -- (12,0);
 	\draw[ color=orange, ultra thick] (0.5,-1) -- (3,0);
 	\draw[ color=orange, ultra thick] (2,0) -- (4.5,-1);
 	\draw[ color=orange, ultra thick] (3.5,-1) -- (6,0);
 	\draw[ color=orange, ultra thick] (5,0) -- (7.5,-1);
 	\draw[ color=orange, ultra thick, dashed] (6.5,-1) -- (8.37501,-0.249996);
 	\draw[ color=orange, ultra thick, dashed] (8,0) -- (8.37501,-0.150004);
 	\draw[ color=orange, ultra thick,dashed] (9.8,-0.32) -- (11.5,-1);
 	\draw[color=red, ultra thick] (0,0) -- (2,0);
 	\draw[ color=black, ultra thick] (2,0) -- (3,0);
 	\draw[color=red, ultra thick] (3,0) -- (5,0);
 	\draw[color=black, ultra thick] (5,0) -- (6,0);
 	\draw[color=red, ultra thick] (6,0) -- (8,0);
 	\draw[color=black, ultra thick, dashed] (8,0) -- (8.37501,0);
 	\draw[color=black, ultra thick] (0.5,-1) -- (1.5,-1);
 	\draw[color=red, ultra thick] (1.5,-1) -- (3.5,-1);
 	\draw[color=black, ultra thick] (3.5,-1) -- (4.5,-1);
 	\draw[color=red, ultra thick] (4.7,-1) -- (6.5,-1);
 	\draw[color=black, ultra thick] (6.5,-1) -- (7.5,-1);
 	\draw[color=red, ultra thick, dashed] (7.5,-1) -- (8.37501,-1);
 	\draw[color=black, ultra thick] (10.5,-1) -- (11.5,-1);
 	\draw[color=red, ultra thick, dashed] (11,0) -- (12,0);
 	\shade[shading=ball, ball color=blue] (0,0) circle (.2);
 	\shade[shading=ball, ball color=blue] (2,0) circle (.2);
 	\shade[shading=ball, ball color=blue] (3.0,0) circle (.2);
 	\shade[shading=ball, ball color=blue] (5,0) circle (.2);
 	\shade[shading=ball, ball color=blue] (6,0) circle (.2);
 	\shade[shading=ball, ball color=blue] (8,0) circle (.2);
 	\shade[shading=ball, ball color=blue] (12,0) circle (.2);
 	\shade[shading=ball, ball color=blue] (0.5,-1) circle (.2);
 	\shade[shading=ball, ball color=blue] (1.5,-1) circle (.2);
 	\shade[shading=ball, ball color=blue] (3.5,-1) circle (.2);
 	\shade[shading=ball, ball color=blue] (4.5,-1) circle (.2);
 	\shade[shading=ball, ball color=blue] (6.5,-1) circle (.2);
 	\shade[shading=ball, ball color=blue] (7.5,-1) circle (.2);
 	\shade[shading=ball, ball color=blue] (10.5,-1) circle (.2);
 	\shade[shading=ball, ball color=blue] (11.5,-1) circle (.2);
 	\draw[black, ultra thick, dotted] (8.83751,-0.5) -- (9.33751,-0.5);
 	\end{tikzpicture}
 	}
\caption{Sketch of the Kitaev chain extended to $r=2$ nearest neighbor hopping and pairing. The dots represents the Majorana operators ordered from left to right (shifted vertically in order to better draw the links), while the links 
represent the couplings: the green lines 
represent $\mu$, the black lines $(\Delta_1+w_1)$, the red lines 
$(\Delta_1-w_1)$, the orange lines $(\Delta_2+w_2)$, the gray lines $(\Delta_2-w_2)$.}
        \label{fig:2neighborsH}
\end{figure}

\begin{figure}
 	\resizebox{\columnwidth}{!}{
 	\begin{tikzpicture}
 	\draw[ color=orange, ultra thick] (0.5,-1) -- (3,0);
 	\draw[ color=orange, ultra thick] (2,0) -- (4.5,-1);
 	\draw[ color=orange, ultra thick] (3.5,-1) -- (6,0);
 	\draw[ color=orange, ultra thick] (5,0) -- (7.5,-1);
 	\draw[ color=orange, ultra thick, dashed] (6.5,-1) -- (8.37501,-0.249996);
 	\draw[ color=orange, ultra thick, dashed] (8,0) -- (8.37501,-0.150004);
 	\draw[ color=orange, ultra thick,dashed] (9.8,-0.32) -- (11.5,-1);
 	\draw[ color=black, ultra thick] (2,0) -- (3,0);
 	\draw[color=black, ultra thick] (5,0) -- (6,0);
 	\draw[color=black, ultra thick, dashed] (8,0) -- (8.37501,0);
 	\draw[color=black, ultra thick] (0.5,-1) -- (1.5,-1);
 	\draw[color=black, ultra thick] (3.5,-1) -- (4.5,-1);
 	\draw[color=black, ultra thick] (6.5,-1) -- (7.5,-1);
 	\draw[color=black, ultra thick] (10.5,-1) -- (11.5,-1);
 	\shade[shading=ball, ball color=gray] (0,0) circle (.2);
 	\shade[shading=ball, ball color=blue] (2,0) circle (.2);
 	\shade[shading=ball, ball color=blue] (3.0,0) circle (.2);
 	\shade[shading=ball, ball color=blue] (5,0) circle (.2);
 	\shade[shading=ball, ball color=blue] (6,0) circle (.2);
 	\shade[shading=ball, ball color=blue] (8,0) circle (.2);
 	\shade[shading=ball, ball color=gray] (12,0) circle (.2);
 	\shade[shading=ball, ball color=blue] (0.5,-1) circle (.2);
 	\shade[shading=ball, ball color=blue] (1.5,-1) circle (.2);
 	\shade[shading=ball, ball color=blue] (3.5,-1) circle (.2);
 	\shade[shading=ball, ball color=blue] (4.5,-1) circle (.2);
 	\shade[shading=ball, ball color=blue] (6.5,-1) circle (.2);
 	\shade[shading=ball, ball color=blue] (7.5,-1) circle (.2);
 	\shade[shading=ball, ball color=blue] (10.5,-1) circle (.2);
 	\shade[shading=ball, ball color=blue] (11.5,-1) circle (.2);
 	\draw[black, ultra thick, dotted] (8.83751,-0.5) -- (9.33751,-0.5);
 	\end{tikzpicture}
 }
	\caption{Kitaev chain with $r=2$ and $\Delta_1=w_1$, $\Delta_2=w_2$. 
The above system is equivalent to a chain of $2L-2$ Majorana 
fermions interacting with their first neighbors by alternate potentials, 
$2w_1$ and $2w_2$ (described by Eq.~(\ref{Hw}) for $r=2$), plus two unpaired Majorana operators at the ends of the chain.}
 	\label{fig:1MFs2vicini}
 \end{figure}
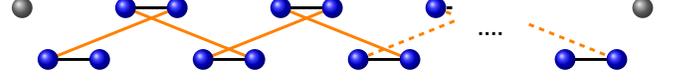

Let us now consider the model Eq.~(\ref{HM}) for the spacial case 
where $\Delta_\ell=w_\ell$ and, 
for simplicity $\mu=0$. In this case the Hamiltonian is simply given by
\begin{equation}
\label{Hw}
H=i\sum_{\ell=1}^{r}\change{\sum_{j=1}^{L-\ell}}
w_\ell\,c_{2j}c_{2(j+\ell)-1}
\end{equation}
The full model in Eq.~(\ref{HM}) can be sketched as in Fig.~\ref{fig:2neighborsH} 
where the case with $r=2$ as been shown.
Some of the couplings in Eq.~(\ref{HM}) disappear for $\mu=0$ and 
$\Delta_\ell=\pm w_\ell$.\\
For $\Delta_\ell= w_\ell$ the Hamiltonian 
Eq.~(\ref{Hw}) can be represented as in Fig.~\ref{fig:1MFs2vicini} where the simple case with $r=2$ is depicted.  
In Eq.~(\ref{Hw}) two Majorana modes are missing, $c_{1}$ and $c_{2L}$, therefore we have one unpaired Majorana mode at each edge. This is consistent with winding number ${\sf w}=1$, Eq.~(\ref{w1}) at $\mu=0$.   
Analogously, for $\Delta_\ell=-w_\ell$, the corresponding Hamiltonian would be 
$H=-i\sum_{\ell=1}^{r}\change{\sum_{j=1}^{L-\ell}}w_\ell\,c_{2j-1}c_{2(j+\ell)}$, in this case the unpaired Majorana operators, the ones which are missing in the Hamiltonian, are $c_2$ and $c_{2L-1}$. 

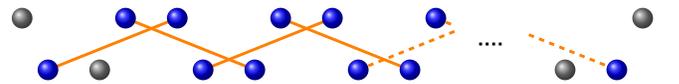
\begin{figure}
 	\resizebox{\columnwidth}{!}{
 	\begin{tikzpicture}
 	\draw[ color=orange, ultra thick] (0.5,-1) -- (3,0);
 	\draw[ color=orange, ultra thick] (2,0) -- (4.5,-1);
 	\draw[ color=orange, ultra thick] (3.5,-1) -- (6,0);
 	\draw[ color=orange, ultra thick] (5,0) -- (7.5,-1);
 	\draw[ color=orange, ultra thick, dashed] (6.5,-1) -- (8.37501,-0.249996);
 	\draw[ color=orange, ultra thick, dashed] (8,0) -- (8.37501,-0.150004);
 	\draw[ color=orange, ultra thick,dashed] (9.8,-0.32) -- (11.5,-1);
 	\shade[shading=ball, ball color=gray] (0,0) circle (.2);
 	\shade[shading=ball, ball color=blue] (2,0) circle (.2);
 	\shade[shading=ball, ball color=blue] (3.0,0) circle (.2);
 	\shade[shading=ball, ball color=blue] (5,0) circle (.2);
 	\shade[shading=ball, ball color=blue] (6,0) circle (.2);
 	\shade[shading=ball, ball color=blue] (8,0) circle (.2);
 	\shade[shading=ball, ball color=gray] (12,0) circle (.2);
 	\shade[shading=ball, ball color=blue] (0.5,-1) circle (.2);
 	\shade[shading=ball, ball color=gray] (1.5,-1) circle (.2);
 	\shade[shading=ball, ball color=blue] (3.5,-1) circle (.2);
 	\shade[shading=ball, ball color=blue] (4.5,-1) circle (.2);
 	\shade[shading=ball, ball color=blue] (6.5,-1) circle (.2);
 	\shade[shading=ball, ball color=blue] (7.5,-1) circle (.2);
 	\shade[shading=ball, ball color=gray] (10.5,-1) circle (.2);
 	\shade[shading=ball, ball color=blue] (11.5,-1) circle (.2);
 	\draw[black, ultra thick, dotted] (8.83751,-0.5) -- (9.33751,-0.5);
 	\end{tikzpicture}
 	 }
 	\caption{Kitaev chain with $r=2$ where $\Delta_1=w_1=\mu=0$ and 
$\Delta_2=w_2\neq 0$, described by the Hamiltonian in Eq.~(\ref{M2vicini}), and four unpaired Majorana operators.} 
\label{fig:2MFs}
 \end{figure}

Let us now consider the special case where $\Delta_r=w_r\neq 0$ and $\Delta_\ell=w_\ell=0, \;{\textrm {for}}\; \ell=1,\dots,r-1$.  
 In this case the Hamiltonian Eq.~(\ref{Hw}) reduces to 
 \begin{equation}
 H=i\sum_{j=1}^{L-r}w_r\,c_{2j}c_{2(j+r)-1}
\label{M2vicini}
 \end{equation}
and we have $2r$ unpaired Majorana operators, $r$ at each edge, $c_{2j-1}$ 
with $j=1,\dots,r$ on the left side and $c_{2(L-j+1)}$ with $j=1,\dots,r$ on 
the right side. An example with $r=2$ is sketched in Fig.~\ref{fig:2MFs}.   
Defining the fermionic operators
 \begin{equation}
\tilde{a}_j=\frac{1}{2}\left(c_{2j}+ic_{2(j+r)+1}\right),\;\;\;
 \tilde{a}_j^{\dagger}=\frac{1}{2}\left(c_{2j}-ic_{2(j+r)+1}\right)
\end{equation}
for $j=1,\dots,L-r$, Eq.~(\ref{M2vicini}) can be written as
\begin{equation}\label{D2vicini}
 H=w_r\sum_{j=1}^{L-r}
\left(2\,\tilde{a}^{\dagger}_j\tilde{a}_j-1
\right)
 \end{equation}
where $r$ highly non local fermions are missing, which can be defined as
\begin{eqnarray}
\tilde{a}_{L-r+j}=\frac{1}{2}\left(c_{2j-1}+ic_{2(L-j+1)}\right)\\
 \tilde{a}_{L-r+j}^{\dagger}=\frac{1}{2}\left(c_{2j-1}-ic_{2(L-j+1)}\right)
\end{eqnarray} 
with $j=1,\dots,r$. Let us define the following parity operator
\begin{eqnarray}
\nonumber P&=&\prod_{j=1}^r P_{L-j+1}
=(-i)^r\prod_{j=1}^r c_{2j-1}c_{2(L-j+1)}\\
&=&\prod_{j=1}^r\left(1-2\,\tilde{a}_{L-r+j}^{\dagger}\tilde{a}_{L-r+j}\right) 
\end{eqnarray}
Since these terms are missing in the Hamiltonian, then $[P,H]=0$, therefore 
$H$ and $P$ have the same eigenstates. Taking the states $|\pm\rangle$ 
such that $\tilde{a}_{j}|+\rangle_j=0$, $\tilde{a}_{j}|-\rangle_j=|+\rangle_j$ 
and $\tilde{a}_{j}^\dagger|+\rangle=|-\rangle_j$, defining  
$|\Psi_0\rangle=\left(\otimes_{j=1}^r|\pm\rangle_{L-r+j}\right)$, we have
\begin{eqnarray}
&&P|\Psi_0\rangle=(-1)^p|\Psi_0\rangle\\
&&H|\Psi_0\rangle=E_0|\Psi_0\rangle
\end{eqnarray}
where $p$ is the number of single particle states $|-\rangle$ contained in $\left(\otimes_{j=1}^r|\pm\rangle_{L-r+j}\right)$ and $E_0=w_r(r-L)$ a constant 
which can be arbitrarily shifted to zero. 
Since we can construct $|\Psi_0\rangle$ choosing for each site $(L-r+j)$, $|+\rangle$ or $|-\rangle$, the degeneracy in energy is $2^r$, divided in two parity sectors, $2^{r-1}$ with odd parity and $2^{r-1}$ with even parity.

 \section{Conclusions}
In this paper we studied the Kitaev chain, generalized 
by allowing for a longer range of the coupling terms. We considered 
the system when time reversal symmetry holds or is broken, characterizing the 
phases by topological invariants and by the spectrum properties, looking at 
the ground state energy and energy gaps. We show that when both the hopping 
and the pairing terms are extended to many neighbors, with time reversal symmetry, we can have many Majorana modes localized at the egdes. This finding is also confirmed by the explicit calculation of the edge modes by means of generalized transfer matrix approach useful to solve the proper 
set of Bogoliubov-de Gennes equations. 
The limit of strictly long-range couplings, with infinite neighbors, has also 
been considered, with and 
without time reversal symmetry, finding several phase diagrams, all characterized by the presence of both massless and massive edge modes.  

\acknowledgements
LD thanks SISSA for kind hospitality and acknowledges financial support from University of Padova through the project BIRD 2016. We thank N. Lo Gullo and L. Lepori for useful discussions.

\appendix
\section{Bogoliubov transformation}  
\label{app:bogoliubov}
The Hamiltonian in Eq.~(\ref{hnostra}), in the fermionic representaion and real space, can be generally written in the matrix form  
\begin{equation}
H=\left(
a^{\dagger}_1 \dots a^{\dagger}_L\,a_1 \dots a_L
\right)
{\cal H}\left(
\begin{array}{c}
\vspace{-0.15cm} a_1\\
\vspace{-0.1cm} \vdots\\
a_L\\
\vspace{-0.15cm} a^{\dagger}_1 \\
\vspace{-0.08cm} \vdots\\
a^{\dagger}_L
\end{array}
\right)
\end{equation}
and can be diagonalized by means of a unitary transformation
 \begin{equation}
 U^\dagger {\cal H} U=\textrm{diag}(\epsilon_1,\dots,\epsilon_L,-\epsilon_1,\dots,-\epsilon_L)
 \end{equation}
getting the typical spectrum of a particle-hole symmetric Hamiltonian. 
On the other hand the Hamiltonian can be written in terms of Majorana operators as in Eq.~(\ref{ourMajo}), which can be written quite in general as it follows
\begin{equation}
\label{HA}
H=i\sum_{i,j}c_iA_{ij} c_j
\end{equation}
where $A$ is a real antisymmetric matrix ($A_{ij}^*=A_{ij}=-A_{ji}$) such that 
 \begin{equation}
 WAW^{T}=\textrm{diag}\Bigg(
\begin{pmatrix}0 & \epsilon_1\\-\epsilon_1 & 0\end{pmatrix},\cdots,\begin{pmatrix}0 & \epsilon_L\\-\epsilon_L & 0\end{pmatrix}\Bigg)
 \end{equation}
 where $\{\pm\epsilon_n\}$ are eigenvalues of ${\cal H}$ 
and $W$ is a real orthogonal matrix, $WW^{T}=W^{T}W=\mathbb{1}$ \cite{enciclopedia}.\\
Under the actions of $U$ and $W$ on respectively fermionic and Majorana operators 
 \begin{equation}\label{transformations}
 \begin{pmatrix}
\vspace{-0.15cm} \tilde a_1\\
\vspace{-0.1cm} \vdots\\
\tilde a_L\\
\vspace{-0.15cm} \tilde a^{\dagger}_1 \\
\vspace{-0.08cm} \vdots\\
\tilde a^{\dagger}_L
 \end{pmatrix}=U^{\dagger}\begin{pmatrix}
\vspace{-0.15cm} a_1\\
\vspace{-0.1cm} \vdots\\
a_L\\
\vspace{-0.15cm} a^{\dagger}_1 \\
\vspace{-0.08cm} \vdots\\
a^{\dagger}_L
 \end{pmatrix},
\qquad \begin{pmatrix}
 b_1 \\
 \vdots \\
 b_{2L} \\
 \end{pmatrix}=W\begin{pmatrix}
 c_1 \\
 \vdots \\
 c_{2L} 
 \end{pmatrix}
 \end{equation}
the Hamiltonian $H$ can be written in the following forms
 \begin{equation}\label{canonical}
 H=\sum_{n=1}^L\epsilon_n\left(2\,\tilde{a}^{\dagger}_n\tilde{a}_n-{1}\right)=i\sum_{n=1}^L\epsilon_nb_{2n-1}b_{2n}
 \end{equation}
 Assuming that $U$ and $W$ are canonical transformations we have
 $\left\{\tilde{a}_n^{\dagger},\tilde{a}_m\right\}=\delta_{n,m}$ and $\left\{\tilde{a}_n,\tilde{a}_m\right\}=0, \,\forall\,n,m=1,\dots,L$, together with $\left\{b_n,b_m\right\}=2\delta_{n,m},\,\forall\,n,m=1,\dots,2L$. 
 The two sets of new operators are linked by
 \begin{eqnarray}
\label{canonic1}
&& b_{2n-1}=\tilde{a}_n+\tilde{a}_n^{\dagger}\\
&& b_{2n}=i\left(\tilde{a}_n^{\dagger}-\tilde{a}_n\right)
\label{canonic2}
 \end{eqnarray}
The quadratic Bogoliubov-de Gennes Hamiltonian $H$, written in terms of fermionic operators, can be 
diagonalized by Bogoliubov transformation, which is represented by the matrix $U$. We can write $\tilde{a}_n$ as 
combination of the first set of operators $a_j$ by means of two sets of functions $u_{n,j}$ and $v_{n,j}$
 \begin{equation}
 \tilde{a}_n=\sum_{j=1}^{L}\left(u^{*}_{n,j}a_j+v^{*}_{n,j}a_j^{\dagger}\right)
 \end{equation}
 with $\sum_{j=1}^L\left(|u_{n,j}|^2+|v_{n,j}|^2\right)=1,\,\forall\,n=1,\dots,L$. Thus we can 
write $U^{\dagger}$ in the following way
 \begin{equation}
 U^{\dagger}=\begin{pmatrix}
 u_{1,1}^* & \dots & u_{1,L}^* &  v_{1,1}^* & \dots & v_{1,L}^*\\
 \vdots & \hphantom & \vdots & \vdots & \hphantom & \vdots\\
 u_{L,1}^* & \dots & u_{L,L}^* &  v_{L,1}^* & \dots & v_{L,L}^*\\
 v_{1,1} & \dots & v_{1,L} & u_{1,1} & \dots & u_{1,L}\\
 \vdots & \hphantom & \vdots & \vdots & \hphantom & \vdots\\
 v_{L,1} & \dots & v_{L,L} & u_{L,1} & \dots & u_{L,L}\\
 \end{pmatrix}
 \end{equation}
From Eq.~(\ref{transformations}) and Eqs.~(\ref{canonic1}),~(\ref{canonic2}), we can write 
 \begin{eqnarray}
\label{UW}
 W_{2n-1,2j-1}&=&\frac{1}{2}\left(u_{n,j}+u_{n,j}^*+v_{n,j}+v_{n,j}^*\right)\\
 W_{2n-1,2j}&=&\frac{i}{2}\left(-u_{n,j}+u_{n,j}^*+v_{n,j}-v_{n,j}^*\right)\\
 W_{2n,2j-1}&=&\frac{i}{2}\left(u_{n,j}-u_{n,j}^*+v_{n,j}-v_{n,j}^*\right)\\
 W_{2n,2j}&=&\frac{1}{2}\left(u_{n,j}+u_{n,j}^*-v_{n,j}-v_{n,j}^*\right)
 \end{eqnarray}
 or, viceversa,
 \begin{eqnarray}
\label{WU}
\nonumber u_{n,j}^*&=&\frac{1}{2}\Big(W_{2n-1,2j-1}+W_{2n,2j}\\
&+&i\big(W_{2n,2j-1}-W_{2n-1,2j}\big)\Big)\\
\nonumber v_{n,j}^*&=&\frac{1}{2}\Big(W_{2n-1,2j-1}-W_{2n,2j}\\
&+&i\big(W_{2n,2j-1}+W_{2n-1,2j}\big)\Big)
 \end{eqnarray}
Now using the Heisenberg equations, derived from Eq.~(\ref{canonical}),
\begin{eqnarray}
\left[H,b_{2n}\right]&=&2i\epsilon_n b_{2n-1}\\
\left[H,b_{2n-1}\right]&=&-2i\epsilon_nb_{2n}
\end{eqnarray} 
expressing $b_n$ in terms of $c_j$ and using Eq.~(\ref{HA}), we can write
\begin{eqnarray}
\epsilon_n W_{2n-1,2j-1}&=&2\sum_{i}W_{2n,i}\,A_{2j-1,i}\\
\epsilon_n W_{2n-1,2j}&=&2\sum_{i}W_{2n,i}\,A_{2j,i}\\
\epsilon_n W_{2n,2j-1}&=&2\sum_{i}W_{2n-1,i}\,A_{i,2j-1}\\
\epsilon_n W_{2n,2j}&=&2\sum_{i}W_{2n-1,i}\,A_{i,2j}
\end{eqnarray}
More specifically, using Eq.~(\ref{ourMajo}) we get the following Bogoliubov 
equations of the wavefunctions for the extended Kitaev model
\begin{eqnarray}
\label{W1}
&&\epsilon_n W_{2n-1,2j-1}=-\mu W_{2n,2j}\\
\nonumber &&\hspace{1cm}+\sum_{\ell=1}^r\Big[ w_\ell 
\sin\varphi_\ell\left(W_{2n,2(j-\ell)-1}-W_{2n,2(j+\ell)-1}\right)\\
\nonumber &&\hspace{1cm}+(\Delta_\ell-w_\ell\cos\varphi_\ell)W_{2n,2(j+\ell)}\\
\nonumber&&\hspace{1cm}-(\Delta_\ell+w_\ell\cos\varphi_\ell)W_{2n,2(j-\ell)}\Big]\\
&&\epsilon_n W_{2n-1,2j}=\mu W_{2n,2j-1}\\  
\nonumber &&\hspace{1cm}+\sum_{\ell=1}^r\Big[ w_\ell
\sin\varphi_\ell\left(W_{2n,2(j-\ell)}-W_{2n,2(j+\ell)}\right)\\
\nonumber  &&\hspace{1cm}
-(\Delta_\ell-w_\ell\cos\varphi_\ell)W_{2n,2(j-\ell)-1}\\
\nonumber &&\hspace{1cm}
+(\Delta_\ell+w_\ell\cos\varphi_\ell)W_{2n,2(j+\ell)-1}\Big]\\
&&\epsilon_n W_{2n,2j-1}=\mu W_{2n-1,2j}\\
\nonumber &&\hspace{1cm}+\sum_{\ell=1}^r\Big[ w_\ell
\sin\varphi_\ell\left(W_{2n-1,2(j+\ell)-1}-W_{2n-1,2(j-\ell)-1}\right)\\
\nonumber &&\hspace{1cm}-(\Delta_\ell-w_\ell\cos\varphi_\ell)
W_{2n-1,2(j+\ell)}\\
\nonumber&&\hspace{1cm}+(\Delta_\ell+w_\ell\cos\varphi_\ell)W_{2n-1,2(j-\ell)}
\Big]\\
\label{W4}
&&\epsilon_n W_{2n,2j}=-\mu W_{2n-1,2j-1}\\
\nonumber &&\hspace{1cm}+\sum_{\ell=1}^r\Big[ w_\ell
\sin\varphi_\ell\left(W_{2n-1,2(j+\ell)}-W_{2n-1,2(j-\ell)}\right)\\
\nonumber  &&\hspace{1cm}
+(\Delta_\ell-w_\ell\cos\varphi_\ell)W_{2n-1,2(j-\ell)-1}\\
\nonumber &&\hspace{1cm}
-(\Delta_\ell+w_\ell\cos\varphi_\ell)W_{2n-1,2(j+\ell)-1}\Big]
\end{eqnarray}
If time reversal symmetry holds ($\varphi_\ell=0$) 
then we can choose $u^*_{n,j}=u_{n,j}$ and 
$v^*_{n,j}=v_{n,j}$, therefore $W_{2n,2j-1}=W_{2n-1,2j}=0$ while, calling 
$W_{2n-1,2j-1}$ the wavefunction $\phi_{\epsilon_n,j}$ and $W_{2n,2j}$ the 
wavefunction $\psi_{\epsilon_n,j}$ 
\begin{eqnarray}
&&\phi_{\epsilon_n,j}\equiv W_{2n-1,2j-1}=u_{n,j}+v_{n,j}\\
&&\psi_{\epsilon_n,j}\equiv W_{2n,2j}\phantom{_{-1-.}}=u_{n,j}-v_{n,j}
\end{eqnarray}
Eqs.~(\ref{W1})-(\ref{W4}) reduce to Eqs.~(\ref{phi}), (\ref{psi}).


\begin{thebibliography}{99}

\bibitem{Wen} X.-G. Wen, \emph{Quantum Field Theory and Many Body Systems},
OUP Oxford, September 06 2007
\bibitem{Kitaev}A. Y. Kitaev, 
\emph{Unpaired Majorana fermions in quantum wire}, Phys. Usp. {\bf 44} 131 
(2001).
\bibitem{Chen} L.-J. Lang and S. Chen, 
\emph{Majorana fermions in density-modulated p-wave superconducting wires}, 
Phys. Rev. B \textbf{86}, 205135 (2012)
\bibitem{Ising} Y. Niu, S. B. Chung, C.-H. Hsu, I. Mandal, S. Raghu and S. Chakravarty, \emph{Majorana zero modes in a quantum Ising chain with longer-ranged interactions}, Phys. Rev. B \textbf{85}, 035110 (2012)
\bibitem{DeGottardi2} W. DeGottardi, D. Sen and S. Vishveshwara, 
\emph{Topological phases, Majorana modes and quench dynamics in a spin ladder system}, New J. Phys. \textbf{13} (2011) 065028
\bibitem{Zvyagin} A. A. Zvyagin, 
\emph{Dynamics of the Kitaev chain model under parametric pumping}, 
Phys. Rev. B \textbf{90}, 014507 (2014)
\bibitem{ExperimentalMFs} S. Nadj-Perge, I. K. Drozdov, J. Li, H. Chen, S. Jeon, J. Seo, A. H. MacDonald, B. A. Bernevig, A. Yazdani, 
\emph{Observation of Majorana fermions in ferromagnetic atomic chains on
a superconductor}, 
Science \textbf{346} 6209 602-607 (2014)
\bibitem{Exp2} V. Mourik, K. Zuo, S. M. Frolov, S. R. Plissard, E. P. A. M. Bakkers, L. P. Kouwenhoven, 
\emph{Signatures of Majorana fermions in hybrid superconductor-semiconductor nanowire devices}, Science \textbf{336}  6084, 1003 (2012)
\bibitem{optics} Z. Xiang, T. Yu, W. Zhang, X Hu, J You,  
\emph{Implementing a topological quantum model using a cavity lattice}
Sci. China Phys. Mech. Astron. \textbf{55} 1549 (2012)
\bibitem{Alicea}J. Alicea, Y. Oreg, G. Refael, F. Von Oppen and M. P. A. Fisher, \emph{Non-Abelian statistics and topological quantum information processing in 1D wire networks}, 
Nature Physics \textbf{7}, 412–417 (2011)
\bibitem{Akhmerov} A. R. Akhmerov, 
\emph{Topological quantum computation away from the ground state with Majorana fermions}, 
Phys. Rev. B \textbf{82}, 020509(R) (2010)
\bibitem{Pfeuty} P. Pfeuty, 
\emph{The one-dimensional Ising model with a transverse field}, 
Ann. Phys. \textbf{57}, 79 (1970)
\bibitem{Lieb} E. Lieb, T.  Schultz and D. Mattis, 
\emph{Two soluble models of an antiferromagnetic chain}
Ann. Phys. \textbf{16}, 407 (1961)
\bibitem{DeGottardi}{W. DeGotterdi, M. Thakurathi, S. Vishveshwara and D. Sen, \emph{Majorana Fermions in superconducting wires: effects of longe-range hopping, broken time-reversal symmetry, and potential landscapes}
Phys. Rev. B \textbf{88}, 165111 (2013)}
\bibitem{Ghazaryan} A. Ghazaryan and T. Chakraborty, 
\emph{Long-range Coulomb interaction and Majorana fermions}
Phys. Rev. B \textbf{92}, 115138 (2015)
\bibitem{Lepori}{D. Vodola, L. Lepori, E. Ercolessi, A. V. Gorshkov, G. Pupillo, \emph{Kitaev chain with long-range pairing}, 
 Phys. Rev. Lett. \textbf{113}, 156402, (2014)}
\bibitem{Lepori2}{D. Vodola, L. Lepori, E. Ercolessi, G. Pupillo, \emph{Long-range Ising and Kitaev models: phases, correlations and edge modes}  
New J. Phys. \textbf{18}, (2016) 015001}
\bibitem{delgado}{O. Viyuela, D. Vodola, G. Pupillo, M. A. Delgado, \emph{Topological massive Dirac edge modeds and long-range superconducting Hamiltonians}, 
Phys. Rev. B {\bf 94}, 125121 (2016)}
\bibitem{Lepori3} L. Lepori, L. Dell'Anna, \emph{Long-range topological insulators and weakened bulk-boundary correspondence}, arXiv:1612.08155
\bibitem{Shi} Z. C. Shi, X. Q. Shao and X. X. Yi, 
\emph{Couplings between Majorana bound states mediated by topologically trivial chains}, arXiv:1507.03657v2 
\bibitem{Pientkaprb} F. Pientka, L. Glazman and F. Von Oppen, \emph{Topological superconducting phase in helical Shiba chains}, Phys. Rev. B {\bf 88}, 155420 (2013); 
\emph{Unconventional topological phase transitions in helical Shiba chains}, Phys. Rev. B {\bf 89}, 180505(R) (2014)
\bibitem{Pientka} F. Pientka, Y. Peng, L. Glazman and F. Von Oppen, 
\emph{Topological superconducting phase and Majorana bound states in Shiba chains}, Phys. Scr. {\bf T164}, 014008 (2015) 
\bibitem{pachos} K. Patrick, T. Neupert, J. K. Pachos, \emph{Topological quantum liquids with long-range couplings}, arXiv:1611.00796
\bibitem{classification} A. Altland and M. R. Zirnbauer, 
\emph{Nonstandard symmetry classes in mesoscopic normal-superconducting hybrid structures}, Phys. Rev. B \textbf{55}, 1142 (1997);   
A. Y. Kitaev, 
\emph{Periodic table for topological insulators and superconductors}, 
AIP Conf. Proc. 1134, 22 (2009)
\bibitem{Tewari}S. Tewari and J. D. Sau, 
\emph{Topological invariants for spin-orbit coupled superconductor nanowires}
Phys. Rev. Lett. \textbf{109}, 150408 (2012)
\bibitem{Zinvariant1} L. Santos, Y. Nishida, C. Chamon and C. Mudry, 
\emph{Counting Majorana zero modes in superconductors}, 
Phys. Rev. B \textbf{83}, 104522 (2011)
\bibitem{Budich}{J.G. Budich and E. Ardonne, \emph{Equivalent topological invariants for one-dimensional Majorana wires in symmetry class $D$}, 
Phys. Rev. B \textbf{88} 075419 (2013)}
\bibitem{readegreen}N. Read and D. Green, \emph{Paired states of fermions in two dimensions with breaking of parity and time-reversal symmetries and the fractional quantum Hall effect},                  
 Phys. Rev. B \textbf{61} 10267 (2000)
\bibitem{enciclopedia}{S. Duplij, W. Siegel and J. Bagger,
 \emph{Concise Enciclopedia of Supersymmetry and noncommutative structures in mathematics and physics},
Kluwer Academic Publisher (2004)}

\end{thebibliography}
\end{document}